\documentclass[]{elsarticle}
\usepackage{graphicx,natbib,amssymb}

\def\arcmin{\mbox{$'$}}
\def\arcsec{\mbox{$''$}}

\begin{document}

\begin{frontmatter}

\title{Homogeneous photometry and star counts in the field of 9 Galactic star clusters}

\author[lbl1]{A.F. Seleznev}
\author[lbl2]{G. Carraro}
\ead{gcarraro@eso.org}
\author[lbl3]{E. Costa}
\author[lbl1]{A.V. Loktin}

\address[lbl1]{Astronomical Observatory, Ural State University, Lenin's ave. 51, Ekaterinburg 620083, Russia}

\address[lbl2]{ESO, Alonso de Cordova 3107, Vitacura, Santiago de Chile, Chile}

\address[lbl3]{Departamento de Astronom\'{\i}a, Universidad de Chile, Casilla 36-D, 
Santiago, Chile}

\begin{abstract}
We present homogeneous $V,I$ CCD  photometry of nine stellar fields in the two inner quadrants
of the Galactic plane. The lines-of-view to most of these fields aim in the direction of the very
inner Galaxy, where the Galactic field is very dense, and extinction is high and patchy. Our
nine fields are, according to several catalogs, centred on Galactic star clusters, namely
Trumpler~13, Trumpler~20, Lynga~4, Hogg~19, Lynga~12, Trumpler~25, Trumpler~26, 
Ruprecht~128, and Trumpler~34. Apart from their coordinates, and in some cases additional
basic data (mainly from the 2MASS archive), their properties are poorly known. By means of star
count techniques and field star decontaminated Color-Magnitude diagrams, the nature and size of
these visual over-densities has been established; and, when possible, new cluster fundamental
parameters have been derived. To strengthen our findings, we complement our data-set with JHK$_{s}$
photometry from the 2MASS archive, that we analyze using a suitably defined
Q-parameter.\\
Most clusters are projected towards the Carina-Sagittarium spiral arm.
Because of that, we detect in the Color Magnitude Diagrams 
of most of the other fields several distinctive sequences produced by young population within
the arm.
All the clusters are of intermediate or
old age. The most interesting cases detected by our study are, perhaps, that of Trumpler~20,
which seems to be much older than previously believed, as indicated by its prominent -and double- red clump;
and that of Hogg~19, a previously overlooked old open cluster, 
whose existence in such regions of the Milky Way is puzzling.
\end{abstract}

\begin{keyword}{{\em(Galaxy:)} open clusters and associations: general } 
\end{keyword}

\end{frontmatter}

\begin{table*}
\tabcolsep 0.1truecm
\caption{Basic parameters of the clusters under investigation.
Coordinates are for J2000.0 equinox}
\begin{flushleft}
\begin{tabular}{clccccc}
\hline
\multicolumn{1}{c}{Label} &
\multicolumn{1}{c}{Name} &
\multicolumn{1}{c}{$RA$}  &
\multicolumn{1}{c}{$DEC$}  &
\multicolumn{1}{c}{$l$} &
\multicolumn{1}{c}{$b$} &
\multicolumn{1}{c}{$E(B-V)_{FIRB}$}\\
\hline
& & {\rm $hh:mm:ss$} & {\rm $^{o}$~:~$^{\prime}$~:~$^{\prime\prime}$} & [deg] & [deg]& mag\\
\hline
\smallskip
  1 & Trumpler~13        & 10:23:48 & -60:08:00 & 285.515 & -2.353 &  1.74\\
  2 & Trumpler~20        & 12:39:34 & -60:37:23 & 301.475 & +2.221 &  1.10\\
  3 & Lynga~4            & 15:33:19 & -55:14:00 & 324.656 & +0.659 &  6.88\\
  4 & Hogg~19            & 16:28:57 & -49:06:00 & 335.088 & -0.302 &  9.32\\
  5 & Lynga~12           & 16:46:04 & -50:46:00 & 335.695 & -3.463 &  1.06\\
  6 & Trumpler~25        & 17:24:29 & -39:01:00 & 339.156 & -1.774 &  1.74\\
  7 & Trumpler~26        & 17:28:33 & -29:30:00 & 357.524 & +2.840 &  1.46\\
  8 & Ruprecht~128       & 17:44:18 & -34:53:00 & 354.778 & -2.864 &  1.05\\
  9 & Trumpler~34        & 18:39:48 & -08:25:00 &  24.119 & -1.264 &  2.81\\
\hline
\end{tabular}
\end{flushleft} 
\end{table*}

\section{Introduction}

This study is a continuation of our homogeneous photometric survey for neglected
open clusters in the inner Galaxy. The main motivations of this survey are twofold:\\

\noindent
{\bf 1.} We are searching for old and intermediate age clusters inside the solar ring
in order to extend the baseline of the radial abundance gradient in the disk, and 
this way contribute to better understand our galaxy's chemical evolution. In spite
of this being a notoriously difficult to observe region -due to the extreme density
of the Galactic disk field, the presence of the bulge, and the highly variable
extinction- we have been able to unravel several intermediate-age clusters (Carraro
et al. 2005a, 2005b, 2006), which we aim to follow up spectroscopically to measure
their metallicity. The present observations will also allow to study the rate of
cluster formation and dissolution in hostile regions of our galaxy such as the above.\\

\noindent
{\bf 2.} We are looking for young clusters and/or spiral features in order to better
trace the location and extent of the inner Galaxy spiral arms Carina-Sagittarius and
Scutum-Crux, and probe the existence of a more distant arm beyond Carina (see Carraro
\& Costa 2009, Baume et. al 2009).\\

\noindent
In this paper we focus on 9 additional fields centred on catalogued open clusters
(Dias et al. 2002): Trumpler~13, Trumpler~20, Lynga~4, Hogg~19, Lynga~12, 
Trumpler~25, Trumpler~26, Ruprecht~128, and Trumpler~34.  Apart from their
coordinates (listed in Table~1), and in some cases additional basic data (discussed
in Section~3), their properties are poorly known.
In the same Table~1 we report the
reddening in the direction of our targets, as provided by Schlegel et al. (1998).
This represents the extinction all the way to infinity, and is only meant to provide
an indication of the upper value we expect for the reddening.\\

\noindent
The layout of the paper is as follows.  In Sect.~2 we provide details on our
observations and data reduction procedure.  In Sect.~3 we summarize previous
results (if any) for the fields under study.  Sects.~4 and 5 are dedicated to star
counts, necessary for the field star decontamination process, and to determine
the structure and extension of each over-density. To this aim, we make use both
of our data-set and of photometric data from the 2MASS archive (Skrutskie et al.
2006).
In Sect.~6. we discuss the
Color-Magnitude Diagram (CMD) of each over-density, and provide estimates of 
the fundamental parameters of those recognized as star clusters. Finally, in
Sect.~7 we summarize our results. 

\subsection{Observations}

The observations were made with a cassegrain focus CCD Imager attached to the
0.9m telescope\footnote{This telescope is operated by the SMARTS consortium,
{\tt http://http://www.astro.yale.edu/smarts}} at Cerro Tololo Inter-American
Observatory (CTIO). This camera is equipped with a Tektronic 2048$\times$2046 CCD
detector with 24$\mu$ pixels, yielding a nominal scale of 0.396 $\arcsec$/pixel and
a field-of-view (FOV) of $13.5\arcmin \times 13.5\arcmin$. Gain and readout noise
were 1.5 e$^-$/ADU and 3.6 e$^-$, respectively. QE and other detector characteristics
can be found at the dedicated webpage\footnote{http://www.ctio.noao.edu/cfccd/cfccd.html}.\\

The observational material was obtained in two observing runs (April and June, 2006),
summarized in Tables~2 and 3.  Both runs were blessed by photometric conditions and
an average seeing of 1.1$\arcsec$.\\

\begin{table}
\centering
\tabcolsep 0.20truecm
\caption{Log of photometric observations on April 19, 2006.}
\begin{flushleft}
\begin{tabular}{lccc}
\hline
Cluster& Filter & Exp time (sec) & Airmass\\
\hline
Lynga~4      & V & 2x5, 30, 600     & 1.03-1.24\\
             & I & 5, 10, 30, 600   & 1.05-1.27\\
Trumpler~13  & V & 2x5, 30, 600     & 1.14-1.30\\
             & I & 5, 10, 30, 600   & 1.12-1.26\\
Trumpler~20  & V & 2x5, 30, 2x600   & 1.20-1.40\\
             & I & 5, 10, 30, 2x600 & 1.17-1.35\\
\hline
\end{tabular}
\end{flushleft} 
\end{table}

\begin{table}
\centering
\tabcolsep 0.15truecm
\caption{Log of photometric observations on June 27-28, 2006.}
\begin{flushleft}
\begin{tabular}{lcccc}
\hline
Cluster& Date & Filter & Exp time (sec) & Airmass\\
\hline
\smallskip
Hogg~19       &  Jun 27 & V   & 2x5, 2x10, 2x600             & 1.03-1.20\\
              &               & I & 2x5, 2x10, 600           & 1.03-1.20\\
Lynga~12      &               & V & 2X5, 2X10, 30, 600       & 1.15-1.24\\
              &               & I & 2x5, 2x10, 30, 600       & 1.16-1.26\\
Trumpler~25   &               & V & 2x5, 2x10, 30, 600       & 1.09-1.33\\
              &               & I & 2x5, 2x10, 30, 600       & 1.12-1.30\\
Trumpler~26   &  Jun 28 & V   & 3x5, 3x10, 2x600             & 1.03-1.54\\
              &               & I & 3x5, 3x10, 2x600         & 1.08-1.50\\
Ruprecht~128  &               & V & 3x5, 3x10, 2x600         & 1.11-1.64\\
              &               & I & 3x5, 3x10, 2x600         & 1.15-1.60\\
Trumpler~34   &               & V & 3x5, 3x10, 2x600         & 1.03-1.84\\
              &               & I & 3x5, 3x10, 2x600         & 1.08-1.80\\
\hline
\end{tabular}
\end{flushleft} 
\end{table}

The nine areas observed are shown in Fig.~1.  Numbers on the upper-left corners
indicate the cluster label, in agreement with Table~1. North is up and East to the 
left. FOV is 13.5$\arcmin$ on a side. Finders were made from 600 sec $V$-band frames.\\

\begin{figure*}
\resizebox{\hsize}{!}{\includegraphics[width=170mm]{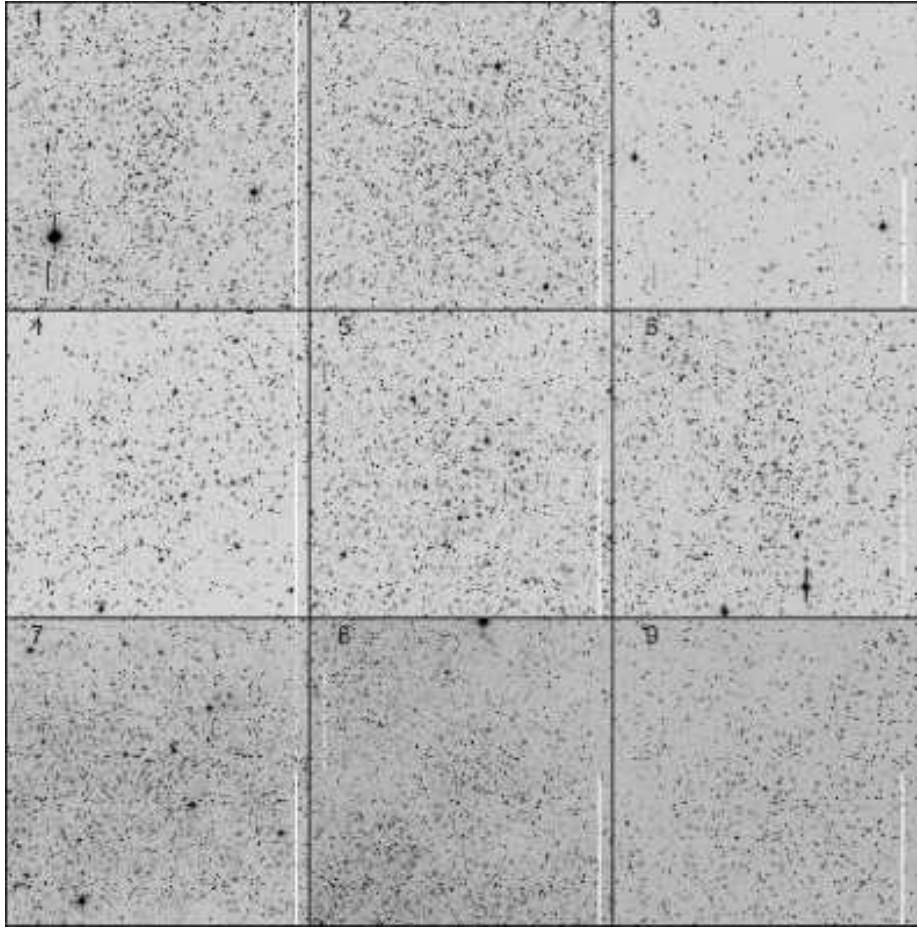}}
\caption{Observed areas . Numbers in the upper-left corners indicate the cluster label,
in agreement with Table~1. North is up and East to the left. FOV is 13.5$\arcmin$ on a
side. Finders were made from 600 sec $V$-band frames.}
\end{figure*}

Our $VRI$ instrumental photometric system was defined by the use of the default
$VRI$ Johnson-Kron-Cousins set available for broad-band photometry on the CTIO 0.9m
telescope. Additional information about them, including their transmission curves,
can be found following this link \footnote{http://www.ctio.noao.edu/instruments/filters/index.html}. \\ 

Five $UBVRI$ standard star areas from the catalog of Landolt (1992) were observed
multiple times each night to determine the transformation of our instrumental
magnitudes to the standard $VRI$ system. A few of the standard areas were followed
each night up to about 2.2 airmasses to optimally determine atmospheric extinction.
Although most of the areas observed include stars of a variety of colors, a few red
standards were observed additionally.

\subsection{Reductions}

Basic calibration of the CCD frames was done using the IRAF\footnote{IRAF is distributed
by the National Optical Astronomy Observatory, which is operated by the Association
of Universities for Research in Astronomy, Inc., under cooperative agreement with
the National Science Foundation.} package CCDRED.  For this purpose, zero-exposure
frames and twilight sky flats were taken every night.  Photometry was performed using
the IRAF DAOPHOT and PHOTCAL packages, and instrumental magnitudes were extracted
following the point spread function (PSF) method (Stetson 1987). The PSF photometry was
aperture-corrected -filter by filter- using aperture corrections determined performing
aperture photometry on a suitable number (typically 15 to 20) of bright stars in the
fields. These corrections were found to vary from 0.08 to 0.21 magnitudes, depending on
filter.

\subsection{The photometry}

In our April 2006 run a grand-total of 187 individual standard star observations
were secured, and we obtained a photometric solution of the form:\\

\noindent
$ v = V + (2.055\pm0.003) + (0.15\pm0.01) \times X + (0.017\pm0.002) \times (V-I)$ \\
$ i = I + (2.945\pm0.002) + (0.06\pm0.01) \times X + (0.027\pm0.002) \times (V-I)$ ,\\
 
Given the very stable photometric conditions encountered in our June 2006 run,
a single photometric solution was derived for all two nights. From a grand-total
of 179 individual standard star observations we obtained:\\

\noindent
$ v = V + (2.079\pm0.003) + (0.16\pm0.01) \times X + (0.024\pm0.002) \times (V-I)$ \\
$ i = I + (2.994\pm0.003) + (0.08\pm0.01) \times X + (0.032\pm0.002) \times (V-I)$ ,\\
 
For both runs, the final {\it r.m.s} of the fitting turned out to be 0.020 and 0.022
for the $V$ and $I$ the pass-bands, respectively.\\

Global photometric errors were estimated using the scheme developed by Patat \& Carraro
(2001, Appendix A1), which takes into account the errors resulting from the PSF fitting
procedure (i.e. from ALLSTAR), and the calibration errors (corresponding to the zero point,
color terms and extinction errors). In Fig.~2 we present global photometric error trends
plotted as a function of $V$ magnitude. Quick inspection shows that most stars brighter than
$V \approx 20$ mag have errors lower than 0.20~mag in magnitude and lower than 0.25~mag in
color. The final photometric catalog will be made available at the WEBDA database
\footnote{http://www.univie.ac.at/webda}.\\

\begin{figure}
\resizebox{\hsize}{!}{\includegraphics[width=\columnwidth]{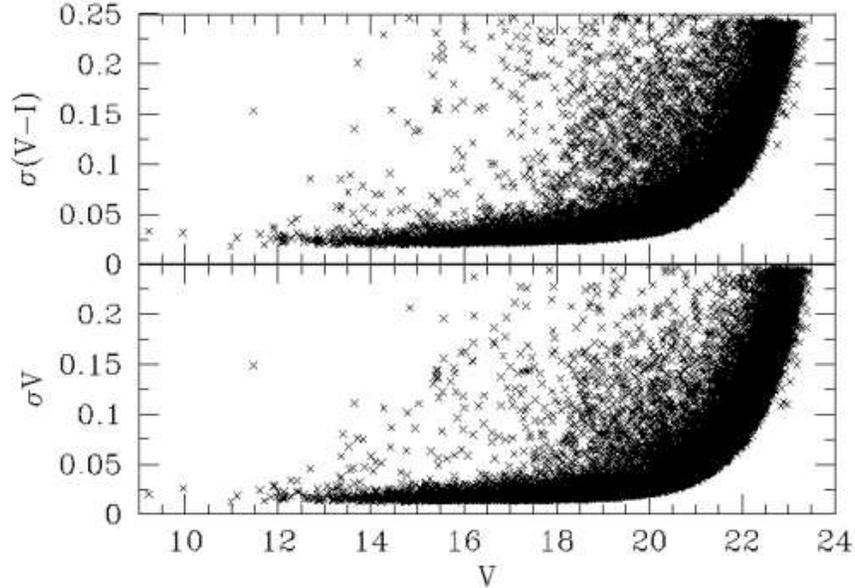}}
\caption{Photometric errors in $V$ and $(V-I)$, plotted as a function of $V$ magnitude}
\end{figure}

\noindent
Completeness corrections were determined by means of artificial-star experiments on
our data.  We created artificial images of each field by adding artificial stars in
random positions to the original images. The artificial stars had the same color and
luminosity distribution as the original sample. In order to avoid the creation of
overcrowding, a maximum of 15$\%$ of the original number of stars was added (between
1000-5000 objects, depending on stellar density). In this way we found that our
completeness level is better than 50$\%$ down to $V$ = 20.5. We have adopted this latter
figure to run our field star decontamination procedure (see Sect.~5).\\

\subsection{Comparison with previous studies}
We compared our photometry with previous studies. The only case for which it was
possible is Trumpler~20 (see next Section), which we compared with Platais et al. (2009).
These authors report BVI photometry of $\sim$ 2500 stars in a 
field of $20^{\prime} \times 20^{\prime}$
centreed on the cluster.
The two studies have different spatial coverage and depth, being our study
deeper but confined to a smaller area.\\
We cross-identified the two photometric catalogues and found 2009 stars in common.
From the comparison we obtain:

\begin{equation}
\Delta V = 0.019 \pm 0.009
\end{equation}

\noindent
and,

\begin{equation}
\Delta (V-I) = 0.024 \pm 0.012.\\
\end{equation}

\noindent
These results show that the two studies agree, since no sizable zero-points offsets
are found.

\begin{figure*}
\resizebox{\hsize}{!}{\includegraphics[width=170mm]{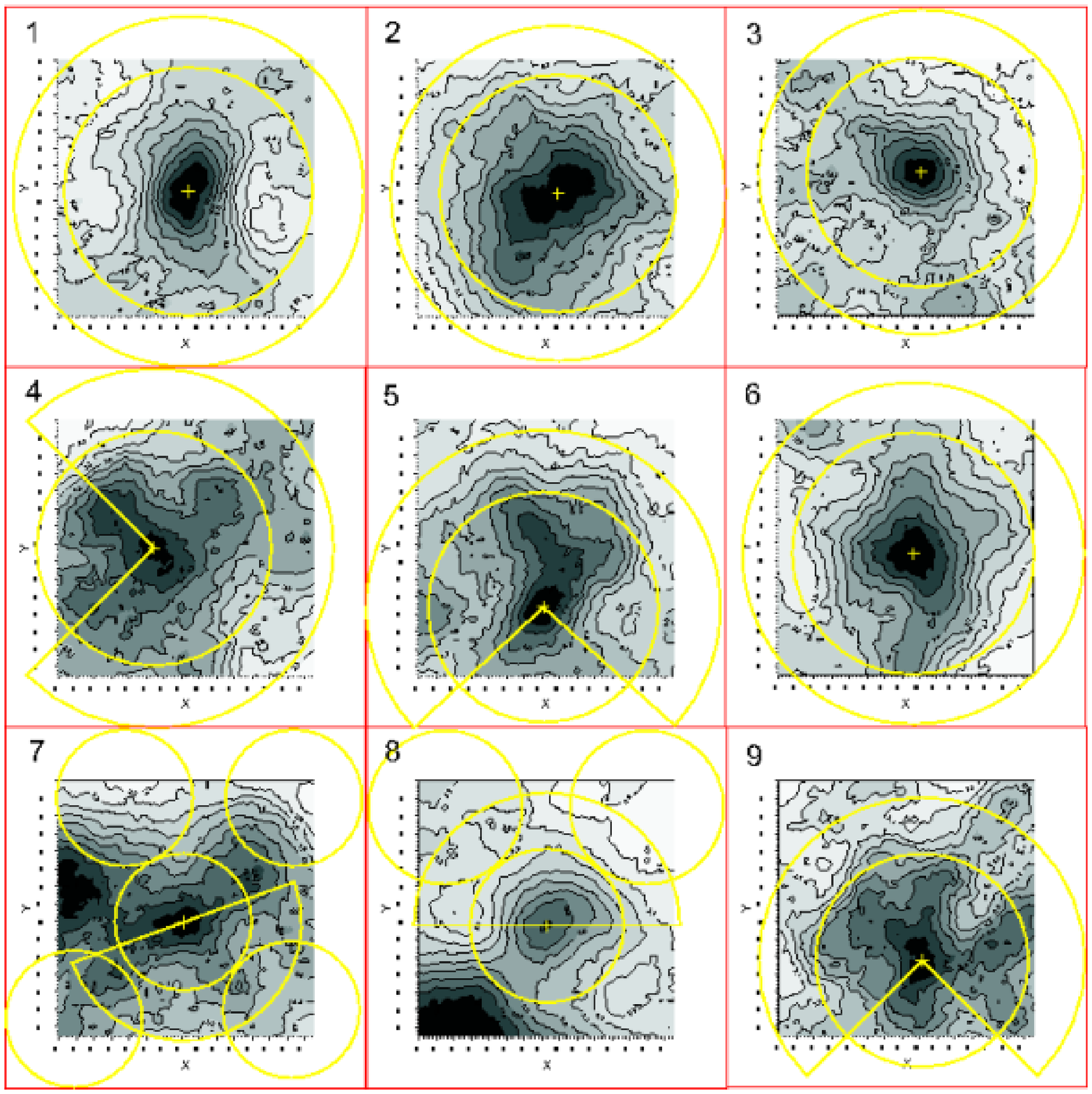}}
\caption{Surface Density Maps for the nine fields under study. Numbers in the upper-left
corners indicate the cluster label, in agreement with Table~1. They have been constructed
using a 300-pixel kernel half-width and a grid of 25-pixel cells. 
Red squares indicate the whole field. See text for more details}
\end{figure*}

\begin{figure*}
\resizebox{\hsize}{!}{\includegraphics[width=170mm]{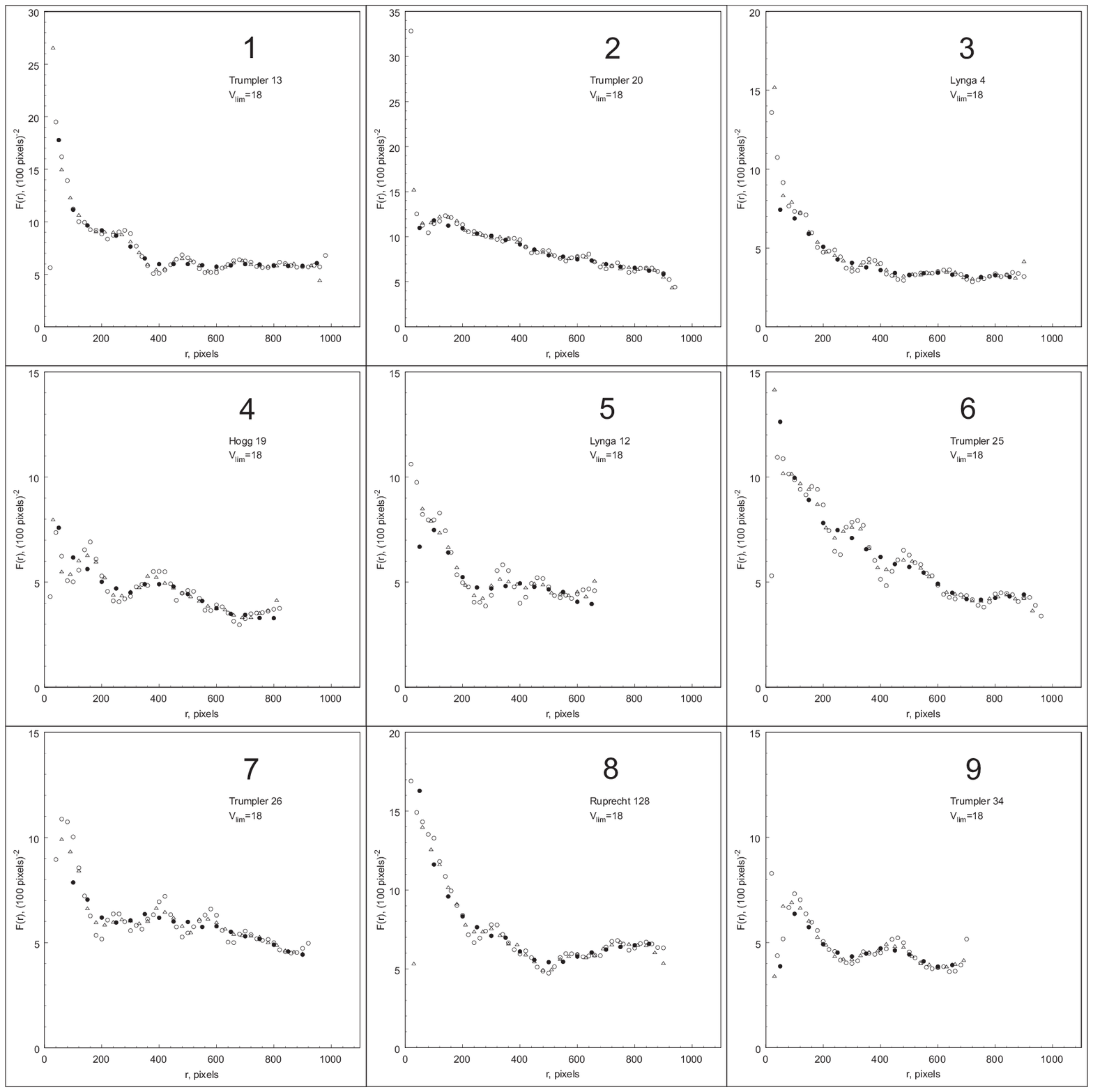}}
\caption{Radial Surface Density Profiles, F(r), for the nine fields under study.  Different
symbols indicate three different increments used to establish N(r): open circles correspond
to 20 pixel (7.38$\arcsec$) steps, open triangles to 30 pixel (11.07$\arcsec$) steps, and
filled circles to 50 pixel (18.45$\arcsec$) steps. See text for details}
\end{figure*}

\section{Previous investigations}

In this Section we summarize previous results, if any, for the fields under investigation.
In most cases we are referring to 2MASS archival data analysis. Each cluster is identified
with its name and the number listed in Table~1.\\

\noindent
{\bf 1. Trumpler~13}\\
Discovered by Trumpler (1930), this object was classified as a medium richness, $\sim5\arcmin$
diameter, star cluster by van den Bergh \& Hagen (1974).  The only observational data for
this object are those given in the 2MASS catalog and discussed by Bica \& Bonatto (2005).
These authors suggest that Trumpler~13 is an intermediate-age cluster ($\sim$ 300 Myr old),
located at 2.5 kpc from the Sun in the third Galactic quadrant. We note that this cluster
is in fact located in the fourth Galactic quadrant (see Table~1).\\

\noindent
{\bf 2. Trumpler~20}\\
This cluster was also discovered by Trumpler (1930), and it is described as a rich open
cluster, with a diameter of $\sim7\arcmin$, by van den Bergh \& Hagen (1974). The only
study of Trumpler~20 that we are aware of is that by McSwain \& Gies (2005), who provide
shallow Stromgren photometry aimed at finding Be stars in open clusters. They suggest that
this cluster is about 150 Myr old, and located at 2.5 kpc from the Sun. Their CMD (their
Fig.~59) shows however a prominent clump, which attracted our attention and seems to
indicate a much larger age. During the revision of this paper we came across to the first
paper on this cluster by Platais et al. (2009), who suggest the cluster is indeed relatively
old basing on optical photometry and Echelle spectroscopy. They derived a reddening E(B-V)=0.46,
an age of 1.3 Gyr and a metallicity [Fe/H]=-0.11. The cluster is found to be located at
3.3 kpc from the Sun.\\

\noindent
{\bf 3. Lynga~4}\\
This cluster is first mentioned in the search for open clusters by Lynga (1964). It has
subsequently been investigated by Moffat \& Vogt (1975), who do not find any indication
for the existence of a cluster at the location of Lynga~4. Humphreys (1976) identified
one supergiant star in the field of Lynga~4 (to which she assigns a distance of 4 kpc),
but did not address the issue of the cluster reality. Recently, from 2MASS photometry,
Bonatto \& Bica (2007) infer that Lynga~4 is indeed a star cluster, of old age
($\sim$1 Gyr), but located at just 1.0 kpc from the Sun.\\

\noindent
{\bf 4. Hogg~19}\\
No studies have been carried out in the field of Hogg~19 after its discovery
by Hogg (1965).\\

\noindent
{\bf 5. Lynga~12}\\
As Lynga~4, this cluster was first mentioned in  Lynga (1964).  The only observational
data-set for this object is that given in the 2MASS catalog and discussed by Bica et al. (2006).
They find that Lynga~12 is a real cluster, at the same distance as Lynga~4 (1.0 kpc), but
with only half the age of the latter.\\

\noindent
{\bf 6. Trumpler~25}\\
Discovered by Trumpler (1930), this object is classified as a medium richness,
$\sim6\arcmin$ diameter, cluster by van den Bergh \& Hagen (1974). To the best of
our knowledge, no other studies have been carried out of the field of this object.\\

\noindent
{\bf 7. Trumpler~26}\\
This cluster was also discovered by Trumpler (1930). The only observational data-set for this
cluster is that given in the 2MASS catalog, and discussed by Bonatto \& Bica (2007). As
was the case of Lynga~4 and Lynga~12, Trumpler~26 also lies at 1 kpc from the Sun. It is
considered to be of intermediate age ($\sim$ 0.7 Gyr).\\

\noindent
{\bf 8. Ruprecht~128}\\
First listed by Ruprecht (1966), this object was subsequently never studied until it was
re-discovered by van den Bergh \& Hagen (1974), and classified as a medium richness
cluster with a diameter of $\sim$ 6 arcmin.\\

\noindent
{\bf 9. Trumpler~34}\\
Discovered by Trumpler (1930). The only study of this cluster is that by McSwain \& Gies
(2005), who provide shallow Stromgren photometry aimed at finding Be stars in open clusters.
They suggest that it is 100 Myr old, and located at 2 kpc from the Sun.\\

\section{Star counts, cluster reality and cluster size}

\subsection{Surface Density Maps and Radial Surface Density Profiles}

Surface Density Maps (SDM) and Radial Surface Density Profiles (RSDP) were constructed 
for all fields under investigation in order to determine each cluster's reality and
size. Example applications of this technique can be found in 
Prisinzano et al. (2001), and Pancino et al.(2003).\\

SDM were constructed using the kernel estimation method (see e.g. Silverman 1986),
with a kernel half-width of 300 pixel (corresponding to 1.845$\arcmin$), and a grid of
25-pixel cells.  The large kernel half-width (HW) chosen is meant to diminish the effect of
density fluctuations (and avoid, for example, numerous density peaks inside a cluster),
and in order to detect the cluster centre clearly. Only stars brighter than $V$=18 mag
were considered because the inclusion of faint stars usually has the negative effect of
making the cluster disappear against the background. To avoid undersampling, we only made
use of the 1450 $\times$ 1450 pixel ($\sim9.6\arcmin\times9.6\arcmin$) central region.
The resulting SDMs are shown in Fig.~3, where the isodensity contour lines plotted are
in units of (100 pixel)$^{-2}$.\\

New, rough coordinates for the clusters centres were obtained from the centre of symmetry of
the inner (maximum) density contours.  The new clusters centres coordinates are given in
Table 4, and they are depicted by crosses in Fig.~3. It should be noted that the use of 
sophisticated methods for cluster centre determination do not make sense in this case, because
the position of the cluster centre clearly depends on limiting magnitude, and on kernel
half-width.\\

Following the procedure described in Seleznev (1994), RSDP, F(r), were obtained by
differentiation of the polynomials fitted to the cumulative star counts function (the number
of stars inside a circle of radius r), N(r).  Third order polynomials were employed in
all cases .\\

The resulting RSDP, for each field are shown in the Fig.~4. Different symbols indicate
three different increments (steps) used to construct N(r): open circles correspond to
20 pixel (7.38$\arcsec$) steps, open triangles to 30 pixel (11.07$\arcsec$) steps, 
and filled circles to 50 pixel (18.45$\arcsec$) steps. F(r) is shown in units of
(100 pixel)$^{-2}$, and only stars brighter than $V$=18 mag were considered, as was the
case of the SDMs shown in Fig.~3.\\

Close inspection of Fig.~4 may draw the attention to the small values of F(r) at the
cluster centres position in some cases. 
They are due to irregularities in the field in some cases caused
by patchy extinction and/or field density fluctuations. 
Technically, the reason is that the cluster centres were determined
from SDMs constructed with a large kernel half-width (300 pixel), whereas the RSDPs have
been derived adopting smaller values for the kernel width. The smaller scale produces a
fluctuating profile, and as a consequence low density values can be obtained at the centres
when N(r) (and therefore F(r)) constructed  using small increments.

\subsection{Star counts}
With the aim of obtaining field star decontaminated CMDs (see Sect. 5), the fields were
divided into inner ({\it cluster}) and outer ({\it comparison field}) regions of equal area
(see Fig.~3). Circular
areas were used as inner regions, and, when possible, full rings were used as outer regions.
When the cluster centre was found to be too close to the field boundary, ring sectors -with
an area equal to that of the corresponding inner circles- were used as outer regions.  In
the case of fields 8 and 9, where there is more than one high density fluctuation, relatively
small circular inner regions, containing only the cluster core, were selected. For these two
fields, the comparison regions used were both ring sectors and circles, equal in area to the
corresponding inner regions.\\

Due to the relatively small size of our fields we cannot use quantitative statistical methods
for cluster size determination (Danilov et al. 1986, Danilov and Seleznev 1994); therefore
we cannot prove that the inner regions completely contain the clusters.  Furthermore, in
some cases the cluster is larger than our FOV, therefore the inner region would only 
contain the cluster core, and, when using the outer regions for comparison, we would be
subtracting stars both from the field and from the cluster halo. This is not a major problem
because we are mostly interested in the CMD's main features, which would still be visible
(note that in these cases the inner region is much denser than the outer region).
Besides, in these regions of the Milky Way extinction is highly variable, and selecting
comparison field too far apart (see Bonatto and Bica 2007) introduces unpredictable
effects in star counts due to reddening variations which are difficult to properly manage.\\

The parameters defining the regions selected in each field are presented in Table 4.
The first and second columns give the clusters labels and names, respectively, in
agreement with Table~1. Columns (3) and (4) provide the new cluster centres in pixels, and
columns (5) and (6) the newly determined coordinates, respectively.  
Columns (7,8) give the radius of the
inner ({\it cluster}) regions in pixels and arcmin, and column (9) the outer radius of the outer
({\it comparison}) region, in pixels, respectively. Columns (10) and (11) list the
starting and ending position angles, $\phi$, for ring sectors in degrees. These
position angles are measured counterclockwise from the south (positive Y direction).
Values of 0$^o$ or 360$^o$ imply that a full ring has been used. Finally, columns (12) and
(13) give the centres of the circular comparison regions used in the case of fields 8
and 9, as explained above.\\

\begin{table*}
\tabcolsep 0.08truecm
\caption{Cluster centres and parameters defining the inner ({\it cluster}) and outer
({\it comparison}) regions in each field. RA and Dec are the newly determined coordinates of
the clusters centres, obtained from the centre of symmetry of the inner (maximum) density
contours.}
\begin{flushleft}
\begin{tabular}{clccccccccccc}
\hline
\multicolumn{1}{c}{Label} &
\multicolumn{1}{c}{Name} &
\multicolumn{1}{c}{$X_C$}  &
\multicolumn{1}{c}{$Y_C$}  &
\multicolumn{1}{c}{$RA$}  &
\multicolumn{1}{c}{$Dec$}  &
\multicolumn{1}{c}{$r_1$} &
\multicolumn{1}{c}{$r_1$} &
\multicolumn{1}{c}{$r_0$} &
\multicolumn{1}{c}{$\phi_1$}&
\multicolumn{1}{c}{$\phi_2$}&
\multicolumn{1}{c}{$X^{'}_C$}&
\multicolumn{1}{c}{$Y^{'}_C$}\\
\hline
& & pixel & pixel & {\rm $hh:mm:ss$} & {\rm $^{o}$~:~$^{\prime}$~:~$^{\prime\prime}$} & pixel & ${\prime}$ & pixel & deg. & deg. & pixel & pixel\\
\hline
\smallskip
1 &  Trumpler 13 &   1040&   1048& 10:23:48 & -60:08:09 & 709  & 4.7 & 1002.7 &     0&  360&&\\
2 &  Trumpler 20 &   1094&   1059& 12:39:32 & -60:37:36 & 675  & 4.5 &  954.6 &     0&  360&&\\
3 &  Lynga 4     &   1110&    937& 15:33:17 & -55:13:28 & 655.5& 4.3 &  927   &     0&  360&&\\
4 &  Hogg 19     &    842&   1028& 16:28:55 & -49:06:02 & 669  & 4.4 & 1021.9 &   315&  225&&\\
5 &  Lynga 12    &   1010&   1362& 16:46:04 & -50:48:03 & 657  & 4.3 & 1009   &    48&  313&&\\
6 &  Trumpler 25 &   1067&   1059& 17:24:28 & -39:01:13 & 689.4& 4.6 &  975   &     0&  360&&\\
7 &  Trumpler 26 &   1009&   1106& 17:28:33 & -29:30:31 & 390  & 2.6 &  675.5 &   290&  110&    387&  1660\\
  &              &       &       & & &   &         &      &    &     &  1626  &  1602\\
  &              &       &       & & &   &         &      &    &     &  1640  &   400\\
  &              &       &       & & &   &         &      &    &     &   675  &   400\\
8 &  Ruprecht 128&   1028&   1124& 17:44:18 & -34:53:37 &  439 & 2.9 &  760.4 &    90&  270&   1598&   450\\
  &              &       &       & & &   &         &      &    &     &   450  &   450\\
9 & Trumpler 34  &   1114&   1324& 18:39:46 & -08:26:14 &  609 & 4.0 &  930.2 &    45&  315&&\\
\hline
\end{tabular}
\end{flushleft} 
\end{table*}

\subsection{Results from the SDM and RSDP analysis}

\noindent
{\bf 1. Trumpler 13}\\
This cluster is clearly elongated in the South-North direction and has a tail in the
South-West direction. It is not seen in the RSDP (Fig.~4) because this profile is a
spherically symmetric approximation which includes the low-density regions to the East
and West. Trumpler~13 shows a clear transition zone (following terminology of Danilov
and Seleznev (1989); also see Seleznev (1994)), from 160 to 400 pixels. Only the outer
boundary of this transition zone is seen in the RSDP; the cluster's halo is not visible
due to field star fluctuations. Nevertheless, our {\it cluster} (inner) region contains
nearly the entire cluster. Taking into account the tail, we estimate that the cluster's
radius is larger than 400 pixels (2.6$\arcmin$).\\

\noindent
{\bf 2. Trumpler 20}\\
This is a large cluster covering nearly the entire field, as clearly seen in both its
SDM and RSDP. The RSDP indicates that the cluster's radius is larger than 950 pixels
(5.8$\arcmin$). The cluster's core is clearly elongated in South-East/North-West direction,
and it is asymmetric. The {\it cluster} region contains only the dense core of
Trumpler 20 (see Tables 4 and 5).\\ 

\noindent
{\bf 3. Lynga 4}\\
This object looks like a small cluster with symmetric core, but with and asymmetric halo
elongated to the North-East. From its SDM and RSDP we estimate that the lower limit of the
cluster's radius is 500 pixels (3.1$\arcmin$), and 480 pixels (3 $\arcmin$), respectively.\\
The cluster is fully contained inside our {\it cluster} region.\\

\noindent
{\bf 4. Hogg 19}\\
This cluster exhibits a very irregular and asymmetric structure.  The RSDP yields a radius
estimate of 680 pixels (4.2$\arcmin$), or possibly larger, which is slightly more than the inner
region we selected. It is difficult to estimate its radius from the SDM, but its SDM seems
to indicate a larger cluster size.\\

\noindent
{\bf 5. Lynga 12}\\
This over-density has highly asymmetric structure. It is difficult to estimate its radius;
the SDM suggests that it might be larger than 700 pixels (4.3$\arcmin$), and
the RSDP indicates that it is larger than 650 pixels (4$\arcmin$). The complicated
structure seen in the SDM may be the result of strong irregularities in the extinction
distribution, giving origin in turn to a cluster-like aspect.\\

\noindent
{\bf 6. Trumpler 25}\\
It is a well-defined cluster with an asymmetric form elongated in the South-North direction.
The SDM does not show the cluster's boundaries, and the RSDP indicates a cluster radius of
more than 960 pixels (5.9$\arcmin$). Our {\it cluster} region contains all of the cluster
core and a large part of its intermediate zone.\\

\noindent
{\bf 7. Trumpler 26}\\
The density maximum considered as the cluster centre seems to be part of a larger structure.
It is not clear if this is a physically connected structure, or a projection effect. It is
very difficult to estimate the cluster radius from the SDM, because it does not have a cluster-like
structure.  The RSDP indicates a small core, and then a gradual decrease of the density outwards.
If it is a true cluster, then its radius is more than 900 pixels (5,5$\arcmin$), and it would
include both the eastern and western density maxima seen in the SDM. \\

\noindent
{\bf 8. Ruprecht 128}\\
This cluster looks like a small fluctuation near a very dense field most probably related to
the Galactic bulge. It is very difficult to estimate the cluster radius from the SDM because it is
overlapped with the density gradient caused by the dense field towards the South-East. The
RSDP gives radius estimate of about 480 pixels (3$\arcmin$), in which case our {\it cluster}
region would include nearly all the cluster.\\

\noindent
{\bf 9. Trumpler 34}\\
The SDM reveals a very irregular and asymmetric structure. The probable cluster centre is
offset with respect to the centre of the field, which makes it  difficult to estimate the
cluster radius from the SDM. The RSDP indicates a cluster radius larger than 600 pixels
(3.7$\arcmin$), while density map suggests an even larger size. We consider this a dubious case,
and will turn back to it in the next Section.\\

\noindent
In Table 5 we summarize our radius estimates obtained from the SDM and RSDP analysis for the
11 clusters studied here.\\

\begin{table}
\tabcolsep 0.35truecm
\caption{Estimates of the cluster's radii, R,  from the SDM and RSDP analysis. A question
mark in the last column indicates a dubious case.}
\begin{flushleft}
\begin{tabular}{clrcc}
\hline
\multicolumn{1}{c}{Label} &
\multicolumn{1}{c}{Name} &
\multicolumn{1}{c}{$R$}  &
\multicolumn{1}{c}{$R$}  &
\multicolumn{1}{c}{$Note$}  \\
\hline
& & pixels & $\arcmin$&\\
\hline
\smallskip
  1 & Trumpler~13        & $\geq$  400 & $\geq$ 2.6&    \\
  2 & Trumpler~20        & $>$   950 & $>$  5.8&    \\
  3 & Lynga~4            & $\geq$  500 & $\geq$ 3.1&    \\
  4 & Hogg~19            & $>$   680 & $>$  4.2&    \\
  5 & Lynga~12           & $\geq$  700 & $\geq$ 4.3& $?$\\
  6 & Trumpler~25        & $>$   960 & $>$  5.9&    \\
  7 & Trumpler~26        & $>$   900 & $>$  5.5& $?$\\
  8 & Ruprecht~128       & $\geq$  480 & $\geq$ 3.0&    \\
  9 & Trumpler~34        & $\geq$  600 & $\geq$ 3.7& $?$\\
\hline
\end{tabular}
\end{flushleft}
\end{table}

\begin{table*}
\tabcolsep 0.35truecm
\caption{Parameters used to analyze 2MASS star counts and revised cluster centres.}
\begin{flushleft}
\begin{tabular}{clrccccc}
\hline
\multicolumn{1}{c}{Label} &
\multicolumn{1}{c}{Name} &
\multicolumn{1}{c}{HW}  &
\multicolumn{1}{c}{Grid}  &
\multicolumn{1}{c}{J$_{lim}$}  &
\multicolumn{1}{c}{(J-H)$_{lim}$}  &
\multicolumn{1}{c}{RA$_{centre}$}  &
\multicolumn{1}{c}{DEC$_{centre}$}  \\
\hline
& &  $\arcmin$ & $\arcmin$ & mag & mag & {\rm $hh:mm:ss$}& {\rm $^{o}$~:~$^{\prime}$~:~$^{\prime\prime}$}\\
\hline
\smallskip
  1 & Trumpler~13        & 3.0 & 0.5 &  16.0 & 0.6 & 10:23:49 & -60:08:12\\
  2 & Trumpler~20        & 5.0 & 0.5 &  16.0 & 0.7 & 12:39:34 & -60:38:42\\
  3 & Lynga~4            & 5.0 & 0.5 &  12.0 & 0.8 & 15:33:23 & -55:14:06\\
  4 & Hogg~19            & 5.0 & 0.5 &  16.0 &     & 16:29:03 & -49:05:24\\
  5 & Lynga~12           & 5.0 & 0.5 &  12.0 & 0.7 & 16:46:06 & -50:45:30\\
  6 & Trumpler~25        & 5.0 & 0.5 &  16.0 & 0.7 & 17:24:30 & -39:00:30\\
  7 & Trumpler~26        & 5.0 & 0.5 &  16.0 & 0.7 & 17:28:35 & -29:28:54\\
  8 & Ruprecht~128       & 3.0 & 0.5 &  16.0 & 0.7 & 17:44:17 & -34:53:06\\
  9 & Trumpler~34        & 5.0 & 0.5 &  14.0 & 0.7 & 18:39:39 & -08:25:48 \\
\hline
\end{tabular}
\end{flushleft}
\end{table*}

\begin{figure}
\resizebox{\hsize}{!}{\includegraphics[width=\columnwidth]{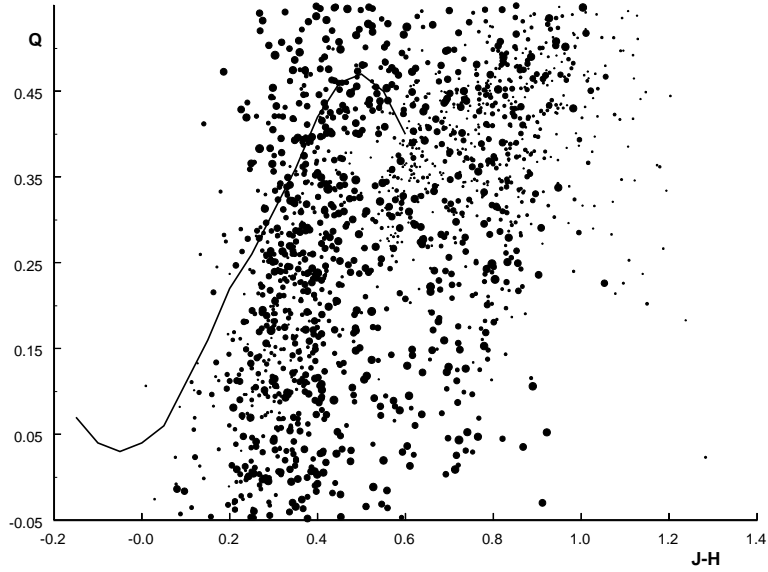}}
\caption{Q-based cluster stars selection for Trumpler 20. The solid line is a reference observational
relation. The size of the dots are proportional to the magnitude errors. See text for details.}
\end{figure}

\section{Results from the 2MASS archival data analysis: star counts and surface density profiles 
in a larger area}

The results of previous Section contain two basic limitations.
Firstly, the star clusters have larger sizes than the area
covered by the detector under use in many cases. Second, in the optical it is more
difficult to account for reddening variations across the clusters' 
field, especially toward the dense inner Galaxy, where we are looking at.\\

\noindent
To cope with these difficulties, we made use of photometry from the
2MASS archive, and re-performed  the star counts analysis in a larger field of view,
to determine in more solid way clusters' reality and radii.\\

\subsection{Cluster members' selection}
In details, we extracted from 2MASS JHK$_{s}$ photometry for stars inside
a box 60 arcmin on a size, and adopted the same technique as in the previous
section to perform star counts, and build up $J$ density maps and clusters'
radial surface density profiles.\\

\noindent
The parameters used and the new cluster centres' coordinated are reported in
Table~6.  The adopted magnitude limits have been chosen to decrease
the noise in star counts and to highlight the cluster more clearly.
Together with a cut-off in magnitude, we also use a color (J-H) cut-off
in the range 0.6$-$0.8 mag, depending on the cluster, to decrease
the amount of expected red field stars.\\
\noindent
An additional, more stringent, criterion has been applied to filter out
interlopers, as follows. 
Firstly, we derived an estimate of the reddening
in the cluster region using the Q vs (J-H) diagram, being Q defined as:

\begin{equation}
Q_{JHK}=(J-H)-\frac{E_{J-H}}{E_{H-K}}\times(H-K),
\end{equation}

\noindent

following Straizys (1992).\\
From Bessell and Brett (1988)we have then:

\begin{equation}
E_{J-H}=0.37 \times E_{B-V}
\end{equation}

\begin{equation}
E_{H-K}=0.19 \times E_{B-V},
\end{equation}

\noindent
and, hence,

\begin{equation}
\frac{E_{J-H}}{E_{H-K}}=0.37/0.19\simeq1.95.
\end{equation}

\noindent
Therefore, we are making use of the following expressions:

\begin{equation}
Q_{JHK}=(J-H)- 1.95 \times(H-K),
\end{equation}

\noindent
and

\begin{equation}
K = K_s +0.044 
\end{equation}

\noindent
from Sarajedini (2004).\\
In details, we started selecting stars in a region
close to the cluster peak (tipycally 5 arcmin), to alleviate field star contamination.
This Q-based selection basically picks up stars having compatible reddening,
and therefore probable clusters' members. This, in turn, results
in a better contrast between cluster and field, and in a more robust estimate
of cluster size and reality.
The method is illustrated in Fig.~5 for the case of Trumpler~20,
one of the most prominent cluster of our sample. In the figure the size of the dots
are proportional to the errors from 2MASS magnitudes, and the solid line
is the above relation calibrated by us with stars from 200 
nearby well studied open clusters.
By shifting horizontally this line we can get 
an estimate of the cluster reddening, which for Trumpler~20 turned
out to be E(J-H) = 0.08.
Then, we extract from the entire
sample all the stars ({\it probable members}) 
having reddening within 0.15 mag from the mean Trumpler~20 reddening. \\

\noindent
We found this procedure effective for Trumpler~13,
Trumpler~20, Hogg~19, Trumpler~25 and Ruprecht~128, for which we
estimated E(J-H) = 0.03, 0.08, 0.16, 0.14, and 0.20,
respectively. In the other 4 cases we could not come out with a
reliable estimate, due to the heavy contamination and noise
of the 2MASS plot. For these latter 4 cases, we used as a first E(J-H) guesses 
estimates from literature data. Namely, we took E(J-H) from Bonatto
\& Bica (2007) for Lynga~4 (0.25) and Trumpler~26 (0.12),
from  Bonatto et al. (2006) for Lynga~12 (0,08), and from
McSwain \& Gies (2005) for Trumpler~34 (0.20).
Adopting these values, we used the Q versus (J-H) diagram
to select stars along the Zero Age Main Sequence (ZAMS), as for
Trumpler~20.  \
These final samples have been used to perform star counts.\\

\begin{figure*}
\resizebox{\hsize}{!}{\includegraphics[width=170mm]{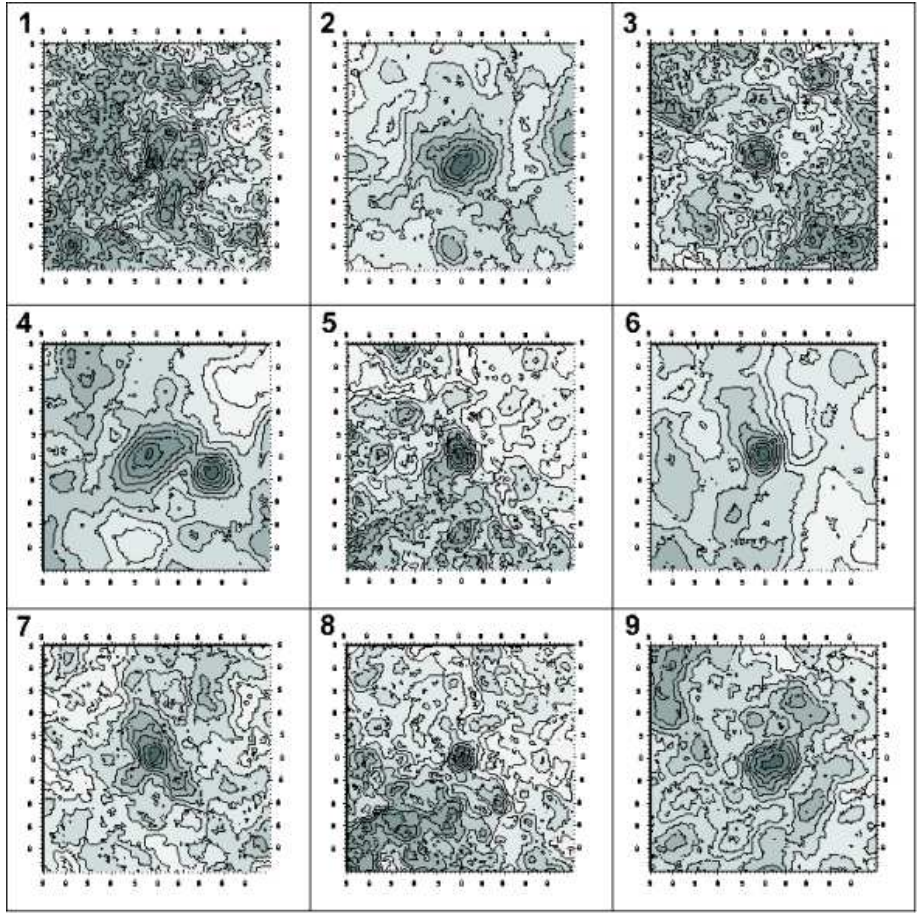}}
\caption{Surface Density Maps from 2MASS for the nine fields under study. Numbers in the upper-left
corners indicate the cluster label, in agreement with Table~1. They have been constructed
using kernel half-widths, magnitude limits in J and grids as in Table~6. In panel~4, the overdensity close
to Hogg~19 is the open cluster NGC~6134}
\end{figure*}

\subsection{Results and comparison with the analysis of the optical data}
We used exactly the same method as for the optical data to perform star counts
and derive radial surface density profiles.
Results are shown in Figs.~ 6 and 7,  and summarized in Tables~ 6 and 7.\\
Table~6 lists the values adopted for the size of the cell grid and half-width
kernel, together with the magnitude and color limits.
The first result is a new determinations of the cluster centers (see columns 7 and 8
in the same table). By comparing these new coordinates with the ones
derived from optical star counts, we find that
there is a general agreement (within less than an arcmin both
in RA and DEC) between the cluster
centres in the optical and in the infra-red, except for Lynga~12, for which
the centre DEC differs by 2.5 arcmin.

\begin{figure*}
\resizebox{\hsize}{!}{\includegraphics[width=170mm]{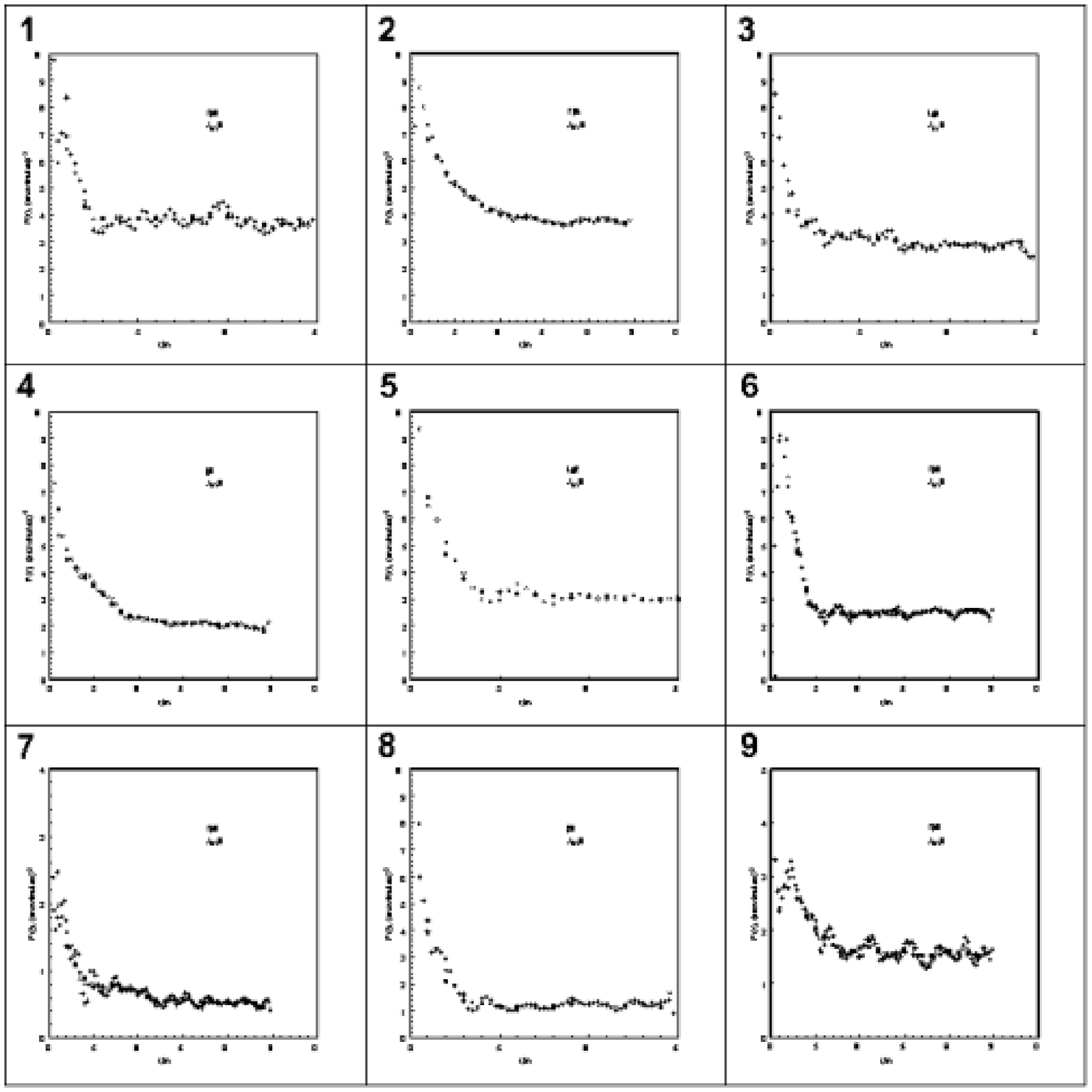}}
\caption{Radial Surface Density Profiles, F(r), for the nine fields under study
as derived from 2MASS. Here filled circles refer to 1 arcmin,
open circlesto 0.5 arcmin, and
crosses to 0.25 arcmin steps.}
\end{figure*}

\begin{table}
\tabcolsep 0.35truecm
\caption{Estimates of the cluster's radii, R,  from the SDM and RSDP analysis.}
\begin{flushleft}
\begin{tabular}{clrc}
\hline
\multicolumn{1}{c}{Label} &
\multicolumn{1}{c}{Name} &
\multicolumn{1}{c}{Radius}  &
\multicolumn{1}{c}{Core radius}  \\
\hline
& & $\arcmin$ & $\arcmin$ \\
\hline
\smallskip
  1 & Trumpler~13        &  3.5 & 2.5 \\
  2 & Trumpler~20        & 17.0 & 5.0 \\
  3 & Lynga~4            &  7.5 & 2.0 \\
  4 & Hogg~19            & 14.0 & 3.0 \\
  5 & Lynga~12           &  8.0 & 4.0 \\
  6 & Trumpler~25        &  7.0 & 4.5 \\
  7 & Trumpler~26        & 13.0 & 4.0 \\
  8 & Ruprecht~128       &  5.0 & 2.0 \\
  9 & Trumpler~34        & 13.0 & 5.0 \\
\hline
\end{tabular}
\end{flushleft}
\end{table}

\noindent
In Table~7 we present new estimates of the clusters' radii (column 3) and core 
radii (column 4). Notice, for the sake of clarity,  that these core radii are not the King core
radii, since we are not fitting any King model (King 1962).
While the clusters' radii we find with 2MASS are, as expected, 
larger than the optical estimates, the core radii we estimate are on the average comparable 
with the adopted cluster area in the optical anlysis (see column 8  -r1- in Table~4).\\
Most cluster stars are presumed to be located inside the core radius,
while outside the core radius still there are cluster stars, but
heavily mixed with the field. 
Basing on that, we are going to
use the core radii from Table~5 to define the clusters' region, and
the regions depicted in Fig.~3 as field regions, to clean in a statistical
way the cluster regions and derive field stars decontaminated
CMDs in the following Section.

\section{Color Magnitude Diagrams and Cluster Fundamental Parameters}

In this Section we make use of the results obtained in previous Sections to construct field star
decontaminated ("clean") CMDs, and to derive new estimates of the clusters fundamental
parameters.

\subsection{Methodology}

To alleviate the high contamination from Galactic disk stars, we employ the same statistical
subtraction technique used in Carraro \& Costa (2007) and in Baume et al. (2007), which was
adapted from Gallart et al. (2003).\\

Briefly, for all objects in the {\it comparison} regions we search for the most similar 
star, in color and magnitude, in the {\it cluster} region, and remove it from the CMD
of the cluster. Matching is done by means of a search ellipse, whose semi-major and semi-minor axis
depend on the photometric errors (see Fig.~2), and their ratio is taken as 5.
If a field star has a counterpart in the cluster area within this ellipse, the
counterpart is removed from the cluster CMD.\\
It should be noted that the procedure also takes into account the completeness level of the
photometry (see Sect. 2). The {\it cluster} region and the {\it comparison} region were
selected as explained in Sect. 4.2.\\

\noindent
Having realized the statistical subtraction, the clean CMDs are compared with theoretical
isochrones from the Padova suite of models (Girardi et al. 2000a). Because we are basically
interested in deriving estimates of the cluster fundamental parameters, which in most cases
are first estimates, adopting the general extinction law is a reasonable assumption. In this
case, the total to selective absorption ratio, $R_V=\frac{A_V}{E(B-V)}$, is equal to 3.1.
As a consequence one can adopt the relation $E(V-I)$ = 1.244 $\times E(B-V)$ to derive $E(B-V)$.
Since we are exploring a region inside the solar ring, adopting a solar metallicity (Z=0.019)
in the models seems to be a reasonable choice. The distance of the Sun from the Galactic centre
was taken as 8.5 kpc, to be homogeneous with our previous studies.\\

\noindent
To assess the reliability of the above procedure, and strenghen our findings, we make also
use of the 2MASS photometry, and build up infrared CMDs. We refer to the 2MASS star counts performed
in the previous section, and extract JHK photometry for all the stars inside the core
radius (see Table~7) and by means of the Q-paramter previouslly described (see Sec.~5). 
This photometry is then analyzed and compared to the same set
of theoretical isochrones. 
We adopt  E(J-H) = 0.29 $\times$ E(V-I) and E(H-K) =0.19$\times$ E(B-V)
from Bessel \& Brett (1988).

\subsection{Cluster Fundamental Parameters}

The results of our analysis are summarized in Table~8, where for each cluster we list the
age, reddening ($E(V-I)$), apparent distance modulus (m-M)$_V$, distance from the
Sun ($d_{\odot}$) and location in the Galactic disk ($X_{GC}$, $Y_{GC}$, $Z_{GC}$, $d_{GC}$).
The uncertainties for the age, reddening and apparent distance modulus given in Table~6 were
derived by adopting different age isochrones (for the sake of the clarity not shown in the
CMDs presented in the next Section), and moving the best fit isochrones back and forth in
the horizontal and vertical direction to adjust reddening and distance modulus.\\
In these series of Figs.~ 8 to 16 we show in the bottom panels, from left to right,
VI photometry for cluster (left panel) and field (middle panel), as selected in Fig.~3 and Table~4,
and the decontaminated CMD (right panel), together with the best fit isochrone.\\

\noindent
This same isochrone is used in the upper panels, where, from the left to the right
we show JHK photometry for the cluster field (left panel, see Table~7), for the stars selected according
to the Q-parameter (middle panel, see  Section~5) and, finally, the CMD with
these latter stars, where an isochrone  fit is provided for the same set of parameters used in the
optical CMD.

\subsection{Color Magnitude Diagrams}
{\bf 1. Trumpler~13}\\
See Fig.~8.
This object is located in the fourth Galactic quadrant just before the tangent to the Carina
branch of the Carina-Sagittarius spiral arm, and for this reason we do not expect important
contamination from spiral features. The CMD of the cluster region differs from that of the
comparison region in the upper part of the Main Sequence (MS). A blue MS, with a turn-off (TO)
at $V \approx 15.5$, is clearly visible in the cluster CMD, but only marginally present in the
comparison CMD, and survives the cleaning process. The blue solar metallicity isochrone plotted
in the right panels is for the fundamental parameters listed in Table~8. Notice the consistency
between the optical and IR results. Apart from the location (in the fourth and not in the third
quadrant) we basically agree with Bica and Bonatto (2005) results.\\

\begin{figure}
\resizebox{\hsize}{!}{\includegraphics[width=\columnwidth]{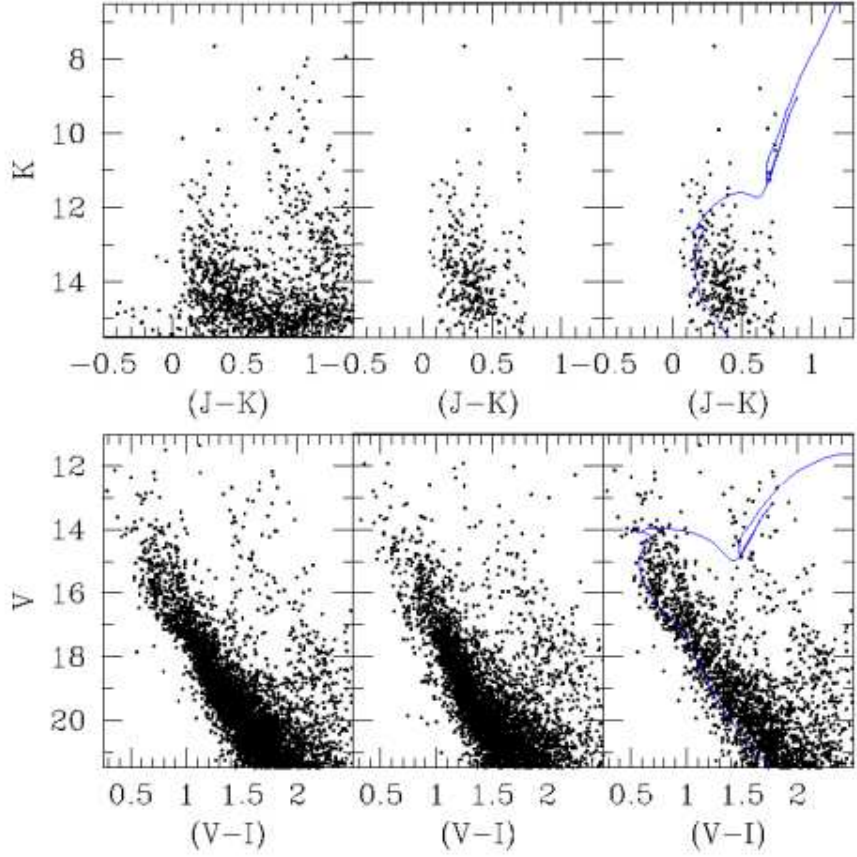}}
\caption{CMDs for Trumpler~13. {\bf Bottom panels}.
In the left panel we show the inner {\it cluster} region CMD,
in the middle panel the outer {\it comparison/field} region CMD, and in the right panel the
corresponding clean CMD (3021 stars). {\bf Top panels}.
In the left panel we show the JHK CMD for the star within the core radius,
in the middle panel the CMD for the same stars but after a selection according
to the Q-parameter. Finally, in the right panels these same latter stars are used
to perform an isochrone fit for set of fundamental
parameters listed in Table~8. The same isochrone is used in the lower right panel.}
\end{figure}

\noindent
{\bf 2. Trumpler~20}\\
See Fig.~9.
Although this cluster is located about 2$^{o}$ above the Galactic plane, some
contamination from young stars of the Carina arm is still visible in the clean CMD.
This sequence was erroneously attributed to Trumpler~20 by McSwain and Gies (2005), but,
by adjusting a Schmidt-Kaler (1982) empirical ZAMS -hereafter empirical ZAMS- (red line),
it can be inferred
that it corresponds to a sector of the Carina arm at a distance of about 2 kpc
(see also Platais et al. 2009, who highlighted the same problem).\\
Trumpler~20 is in fact a much older cluster, as indicated by the conspicuous clump of
red giant branch (RGB) stars seen both in the cluster region CMD, and in the clean CMD.
There is no doubt that Trumpler~20 is an intermediate age cluster, very much resembling
NGC~7789 (Gim et al 1995). It is somewhat surprising that this fact was not noticed before,
and certainly deserves further investigation. The blue solar metallicity isochrone plotted
is consistent with the fundamental parameters listed in Table~8 which, in turn,
nicely agree with the recent study by Platais et al.(2009). Notice the consistency
between the optical and IR results.\\

\noindent
An interesting feature of Trumpler~20 CMD is the presence of a double red clump, which
strenghten its similarity to NGC~7789 and other intermediate age star clusters,
like NGC~5822 and NGC~2660 (Girardi et al. 2000b). 
Such occurence is not limited to star clusters in the Milky Way,
but is also present in the Magellanic Clouds clusters (Girardi et al. 2009).\\

\begin{figure}
\resizebox{\hsize}{!}{\includegraphics[width=\columnwidth]{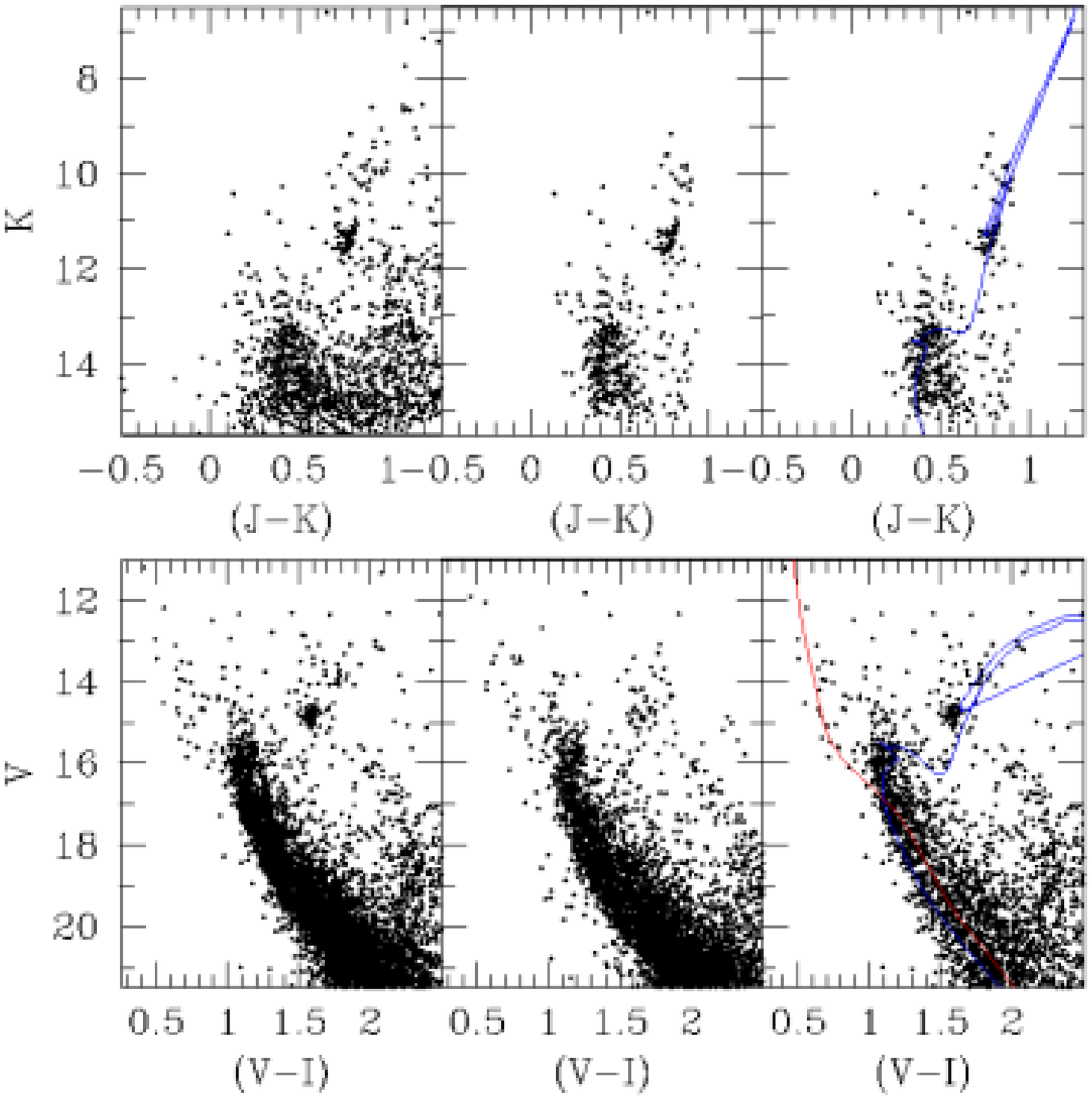}}
\caption{Same as Fig.~8, but for Trumpler~20. The clean optical CMD contains 3250 stars.The blue solar metallicity isochrone is consistent
with the fundamental parameters listed in Table~8, whereas the red line is an empirical ZAMS.}
\end{figure}

\noindent
{\bf 3. Lynga~4}\\
See Fig.~10.
This cluster is clearly visible from IR photometry, and its basic parameters have determined
by fitting the blue isochrones in the upper-right panel. This fit implies an age of 300
million years, a reddening E(V-I)=1.9 and a distance of 1.1 kpc. The age we find is significantly
lower than Bonatto \& Bica (2007) estimate.\\
Due to the extreme absorption,
in the optical CMD (botton panels in Fig.~10) the cluster looks very faint and its MS
is mixed with the general Galactic field stars. However, the bifurcation we see at V$\sim$ 18.0
and (V-I) $\sim$ 1.2, together with the bunch of red stars at 16 $\leq V \leq 17$ (probable giants),
make us confindent about the cluster identification.\\

\begin{figure}
\resizebox{\hsize}{!}{\includegraphics[width=\columnwidth]{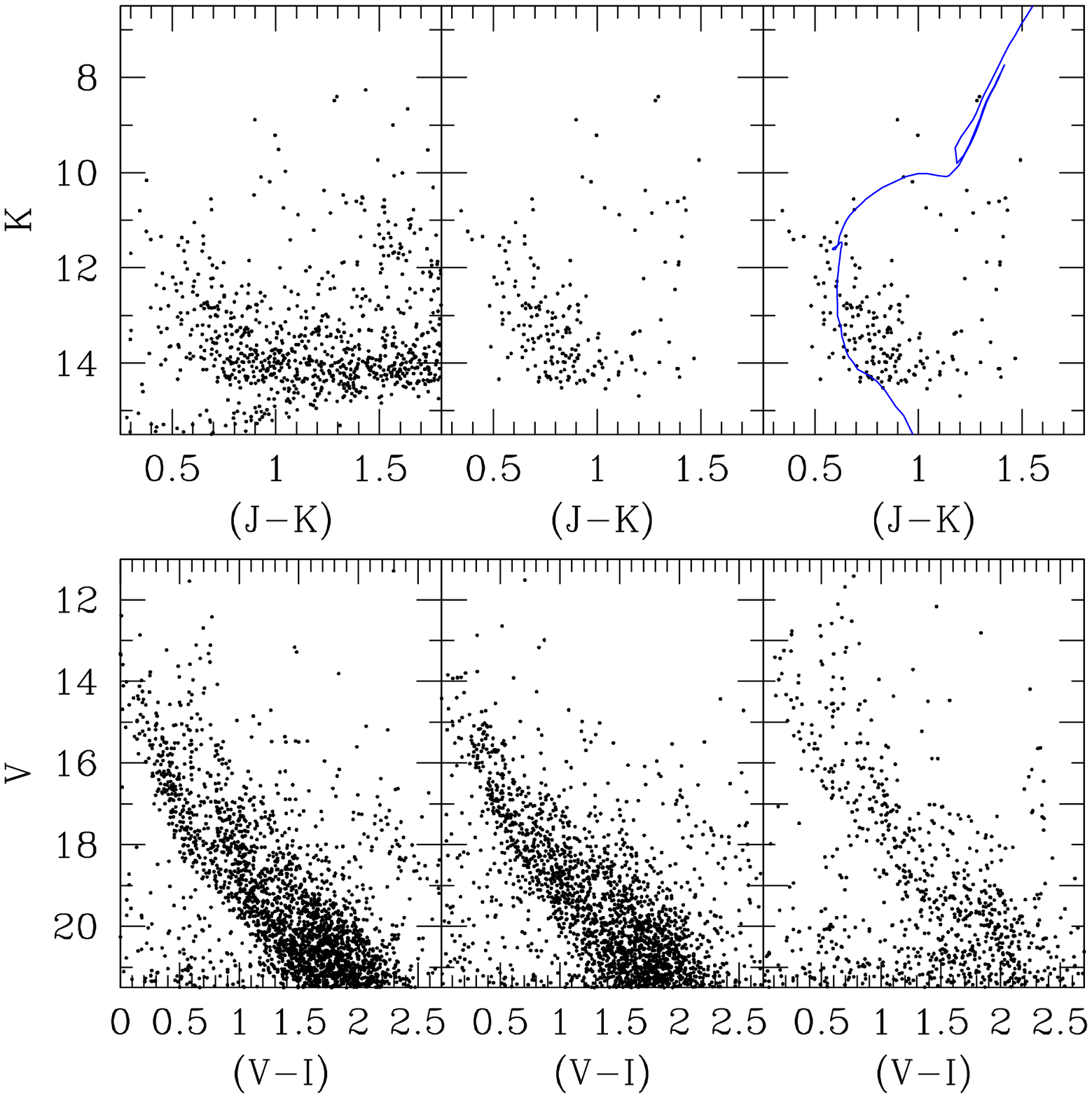}}
\caption{Same as Fig.~8, for Lynga 4. The blue solar metallicity isochrone is consistent
with the fundamental parameters listed in Table~8. We could not see any cluster in the optical
data, and therefore no isochrone fit is shown. The clean optical CMD contains 3185 stars.
See text for details.}
\end{figure}

\noindent
{\bf 4. Hogg~19}\\
See Fig.~11.
This field is located quite low in the Galactic plane (see Table~1), in the direction of
the Carina-Sagittarius spiral arm.  FIRB reddening (Schlegel et al. 1998) in the direction
of Hogg~19, amounts to $\approx$ 21 mag. Three sequences of stars are seen in Fig.~11.
{\bf 1}: a sequence of bright young stars, present both in the cluster region and in the comparison
region, which we interpret as a young diffuse population from the spiral arm;  {\bf 2}: a population
of red giant stars, which is significantly larger in the cluster region than in the field;
{\bf 3}: a fainter thick main sequence, which is much thicker in the cluster region than in the field.
This latter sequence survives in the clean CMD and we relate it to the group of giants stars
that survive as well.  This indicates the presence of an old age star cluster in
the field. We see a Turn Off point at V$\sim$ 18.0 mag and (V-I ) $\sim$ 1.4.
The cluster (Hogg~19) is located in front of the spiral arm. An empirical ZAMS fit
to the young population (red line) yields a distance of 2.4$\pm$0.3 kpc, for a reddening of
0.9$\pm$0.2 mag. The blue solar metallicity isochrone plotted is consistent with the fundamental 
parameters listed in Table~8. An age of about 2  Gyrs is derived both from the optical
and IR data.\\

\begin{figure}
\resizebox{\hsize}{!}{\includegraphics[width=\columnwidth]{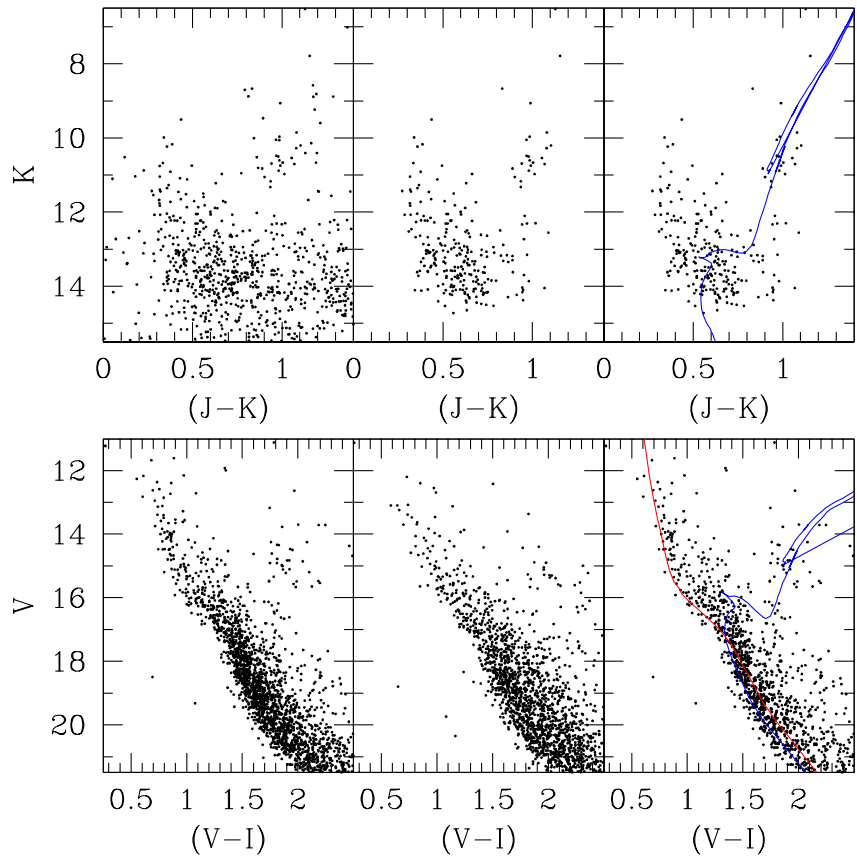}}
\caption{Same as Fig.~8, for Hogg~19. The blue solar metallicity isochrone is consistent with
the fundamental parameters listed in Table~8, whereas the red line is an empirical ZAMS 
drawn to highlight the presence of a strong field contamination caused by the fore-ground
Carina arm. The clean optical CMD contains 1409 stars. See text for details.}

\end{figure}

\noindent
{\bf 5.  Lynga~12}\\
See Fig.~12.
The analysis of 2MASS data reveals that Lynga~12 is a young cluster, suffering heavy extinction.
This is confirmed by our optical data. The fit in the lower right panel of Fig.~12 is with a ZAMS,
shifted by E(V-I) = 1.00 and (m-M) = 13.8, which implies a distance of 1.8 kpc. This young aggregate
is therefore located inside the Carina-Sagittarius spiral arm. It is quite difficult to estimate
the age of the cluster. While in the optical there is no clear indication of evolved stars, IR data
seems to indicate an age around 200 Myr (the red isochrone super-posed in the two right panels).\\

\begin{figure}
\resizebox{\hsize}{!}{\includegraphics[width=\columnwidth]{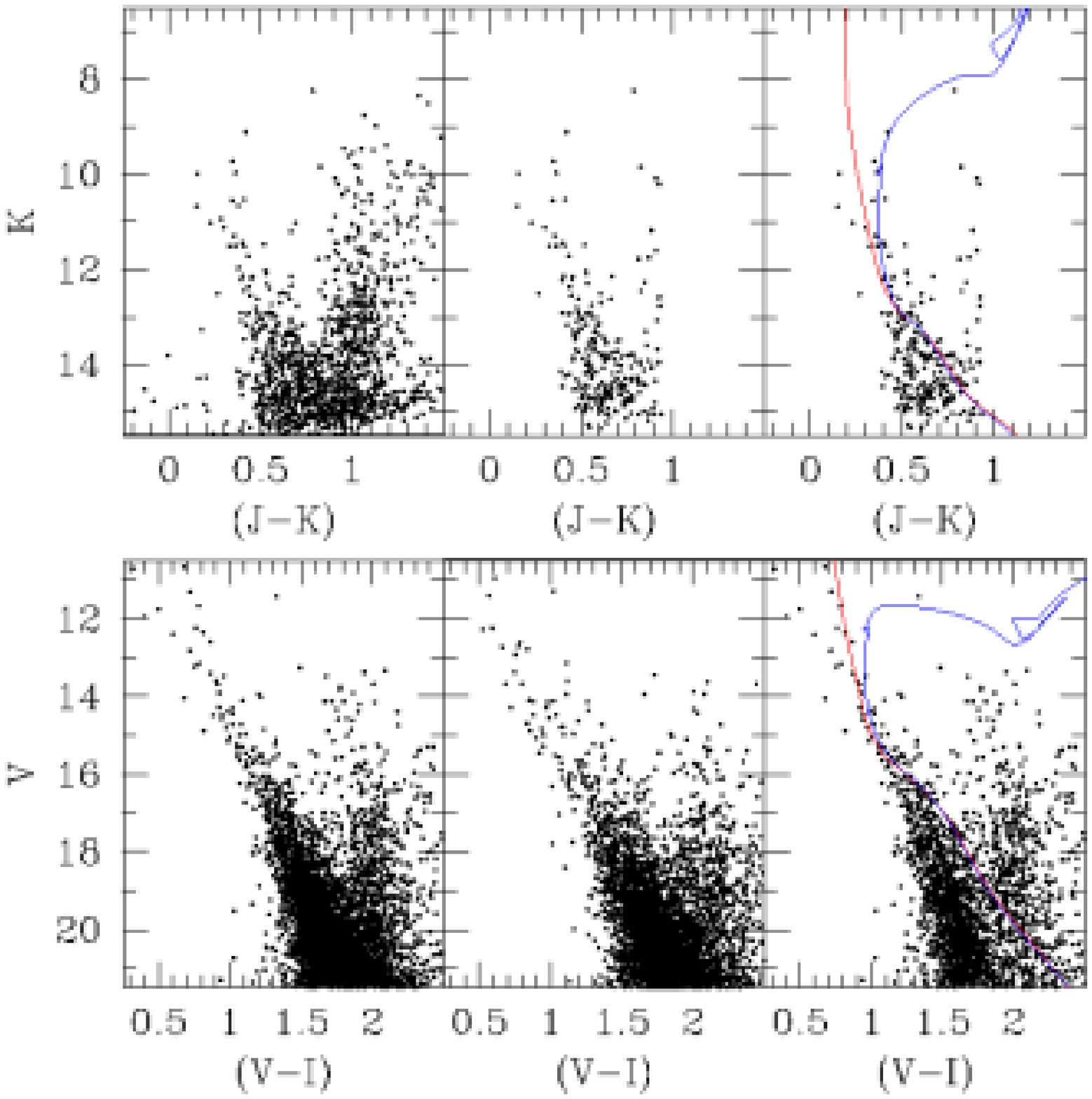}}
\caption{Same as Fig.~8, for Lynga~12.The blue isochrone is consistent with the fundamental
parameters listed in Table~8. The clean optical CMD contains 2894 stars. See text for more details.}
\end{figure}

\noindent
{\bf 6. Trumpler~25}\\
See Fig.~13.
In the cluster region CMD we recognize a bifurcation in the MS at $V \approx 15.5$, together
with an excess of giant stars in comparison to the control field CMD. We tentatively interpret
the bluer MS as a diffuse stellar population in the Carina-Sagittarius arm, while
the red MS is the star cluster Trumpler~25. The isochrone fitted (blue line) indicates 
that this latter is about 0.5 Gyr old, and located at about 2 kpc from the Sun.
Notice the consistency between optical and IR data. 
(see Table~8).\\

\begin{figure}
\resizebox{\hsize}{!}{\includegraphics[width=\columnwidth]{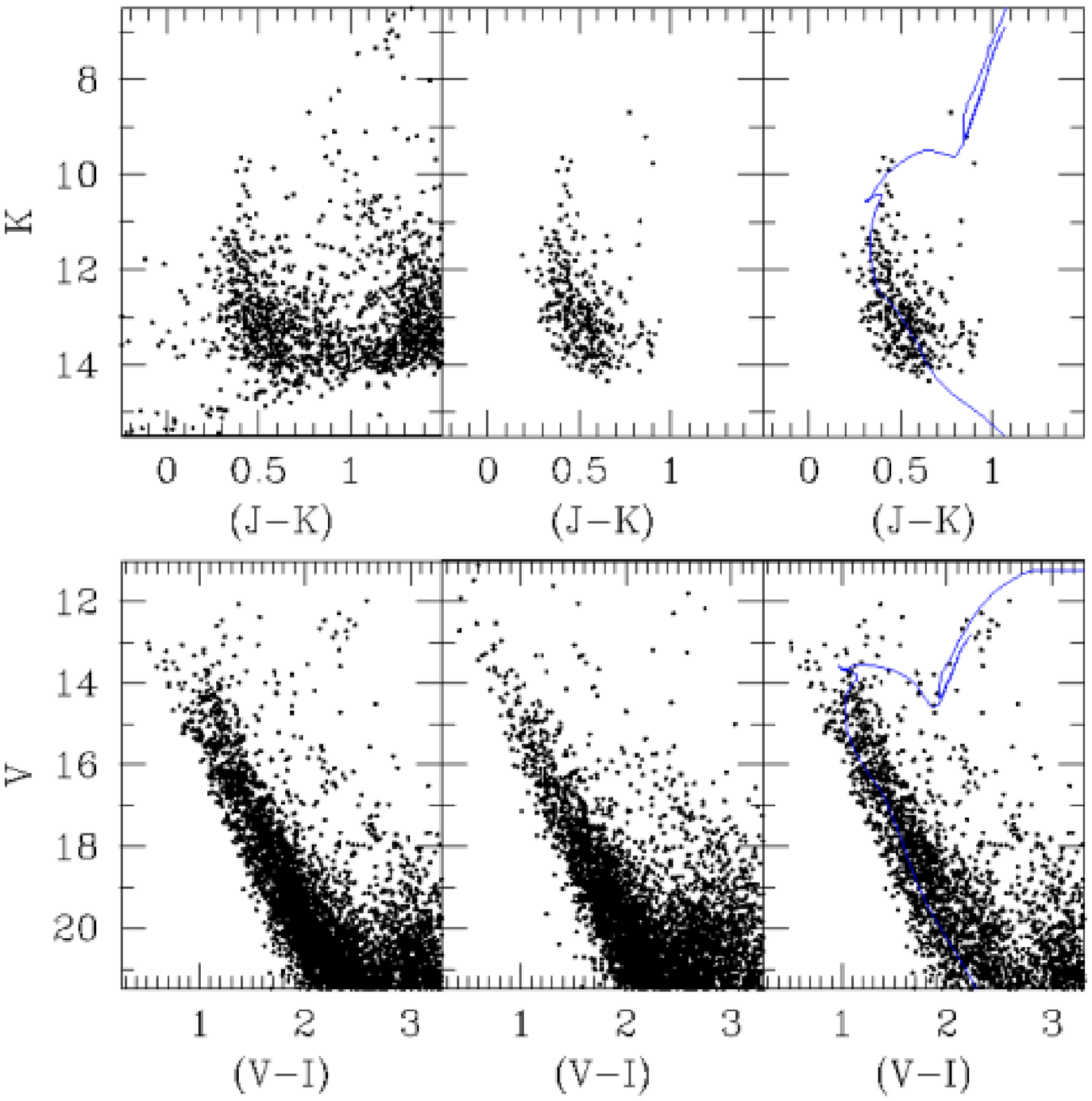}}
\caption{Same as Fig.~8, for Trumpler~25. The blue isochrone is consistent with 
the fundamental parameters listed in Table~8.
The clean optical CMD contains 3030 stars. See text for details.}
\end{figure}

\noindent
{\bf 7. Trumpler~26}\\
See Fig.~14.
For this cluster we defined two {\it comparison} regions (see Fig.~3).  We do not
find any difference by adopting one or the other. This object lies in a direction very close to
the line of sight to the Galactic bulge, which also intersects the Carina-Sagittarius arm. An
examination of Fig.~14 indeed shows a diffuse young stellar population. Both the IR and optical
CMDs provide us with a $\sim$ 300 Myr poorly populated star cluster.
The empirical ZAMS fitted
(blue line) indicates a distance of about 1.22 kpc and a 
reddening E(V-I) = 0.5 mag.\\

\begin{figure}
\resizebox{\hsize}{!}{\includegraphics[width=\columnwidth]{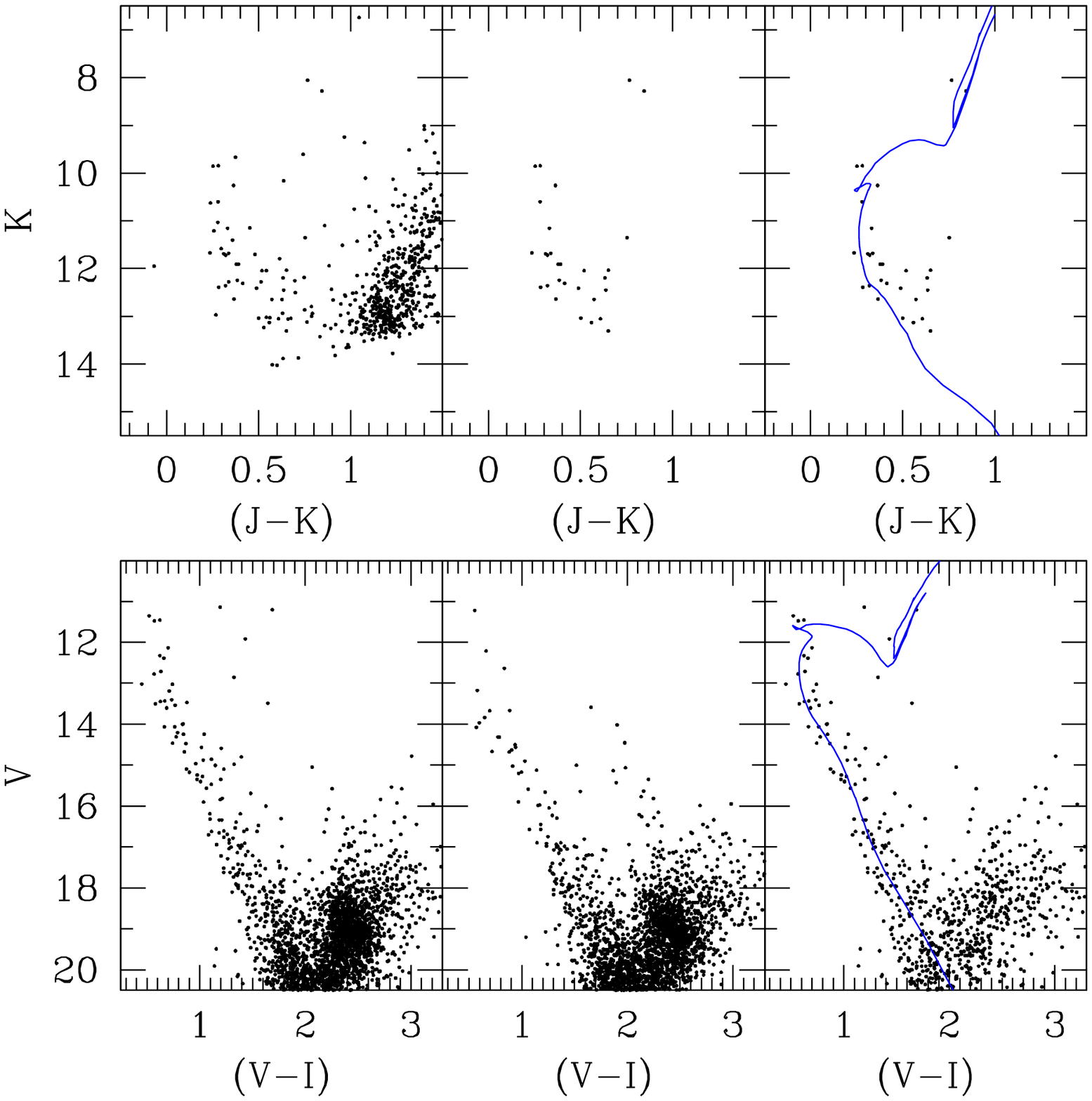}}
\caption{Same as Fig.~8, for Trumpler~26. The blue isochrone is consistent with the fundamental
parameters listed in Table~8. The clean optical CMD contains 878 stars. See text for more details.}
\end{figure}

\begin{table*}
\tabcolsep 0.08truecm
\caption{Derived fundamental parameters of the clusters under investigation.}
\begin{flushleft}
\begin{tabular}{clcccccccc}
\hline
\multicolumn{1}{c}{Label} &
\multicolumn{1}{c}{Name} &
\multicolumn{1}{c}{$Age$}  &
\multicolumn{1}{c}{$E(V-I)$}  &
\multicolumn{1}{c}{$(m-M)_V$} &
\multicolumn{1}{c}{$d_{\odot}$} &
\multicolumn{1}{c}{$X_{GC}$} &
\multicolumn{1}{c}{$Y_{GC}$} &
\multicolumn{1}{c}{$Z_{GC}$} &
\multicolumn{1}{c}{$d_{GC}$} \\
\hline
& & Gyr& mag & mag & kpc& kpc & kpc & kpc & kpc\\
\hline
\smallskip
  1 & Trumpler~13      & 0.4$\pm$0.1 & 0.45$\pm$0.10 & 13.5$\pm$0.2 & 2.9 & 7.8 & -2.8 & -0.1 & 8.3\\
  2 & Trumpler~20      & 1.5$\pm$0.3 & 0.60$\pm$0.10 & 13.9$\pm$0.2 & 3.0 & 6.9 & -2.5 &  0.1 & 7.3\\
  3 & Lynga~4          & 0.3$\pm$0.1 & 1.90$\pm$0.30 & 12.2$\pm$0.2 & 1.1 & 5.3 & -0.6 &  0.0 & 7.6\\
  4 & Hogg~19          & 2.5$\pm$0.3 & 0.80$\pm$0.10 & 14.0$\pm$0.2 & 2.6 & 6.2 & -1.0 &  0.0 & 6.7\\
  5 & Lynga~12         & 0.2$\pm$0.1 & 1.00$\pm$0.10 & 13.8$\pm$0.2 & 1.8 & 6.8 & -0.7 & -0.1 & 6.9\\
  6 & Trumpler~25      & 0.5$\pm$0.1 & 0.90$\pm$0.10 & 13.8$\pm$0.2 & 2.0 & 6.6 & -0.7 &  0.1 & 6.6\\
  7 & Trumpler~26      & 0.3$\pm$0.1 & 0.50$\pm$0.10 & 11.7$\pm$0.2 & 1.2 & 7.3 & -0.0 &  0.0 & 7.3\\
  8 & Ruprecht~128     & 0.8$\pm$0.1 & 1.00$\pm$0.20 & 13.5$\pm$0.2 & 1.6 & 6.9 & -0.1 & -0.1 & 6.9\\
  9 & Trumpler~34      & 0.2$\pm$0.1 & 1.00$\pm$0.10 & 12.9$\pm$0.2 & 1.2 & 7.5 &  0.5 & -0.0 & 7.5\\
\hline
\end{tabular}
\end{flushleft} 
\end{table*}

\noindent
{\bf 8. Ruprecht~128}\\
See Fig.~15.
The situation is similar to that of Trumpler~25. A well defined MS, with a clear TO, typical
of intermediate-age open clusters is seen, together with a few young stars close to a ZAMS.
The isochrone fitted (blue line) indicates an age around 1 Gyr and a heliocentric distance
of 1.6 kpc, for a reddening of about 1 mag (see Table~8). The optical findings are corroborated
by the 2MASS analysis. \\

\begin{figure}
\resizebox{\hsize}{!}{\includegraphics[width=\columnwidth]{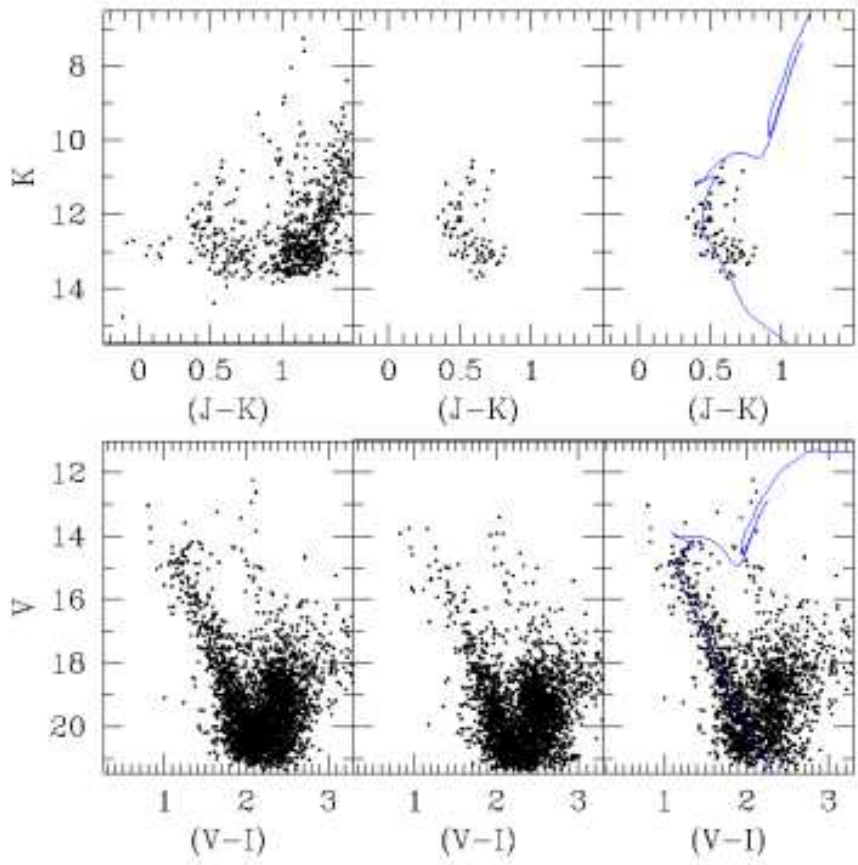}}
\caption{Same as Fig.~8, for Ruprecht~128. The blue isochrone is consistent with the fundamental
parameters listed in Table~8. The clean optical CMD contains 1994 stars. See text for details.}
\end{figure}

\noindent
{\bf 9. Trumpler~34}\\
See Fig.~16.
This cluster is the loosest of the sample, and its density profile
shows it stands weakly above the field and has a hole right in the centre. 
The CMD in the IR is quite broad in color,
and is is difficult to see a clear sequence. Still, we performed some fitting using
the parameters listed in Table~8. The fit is shown by means of the blue
isochrone (right panels of Fig.~16). The cluster turns out to be relatively young,
confirming McSwain \& Gies (2005) suggestions.\\

\begin{figure}
\resizebox{\hsize}{!}{\includegraphics[width=\columnwidth]{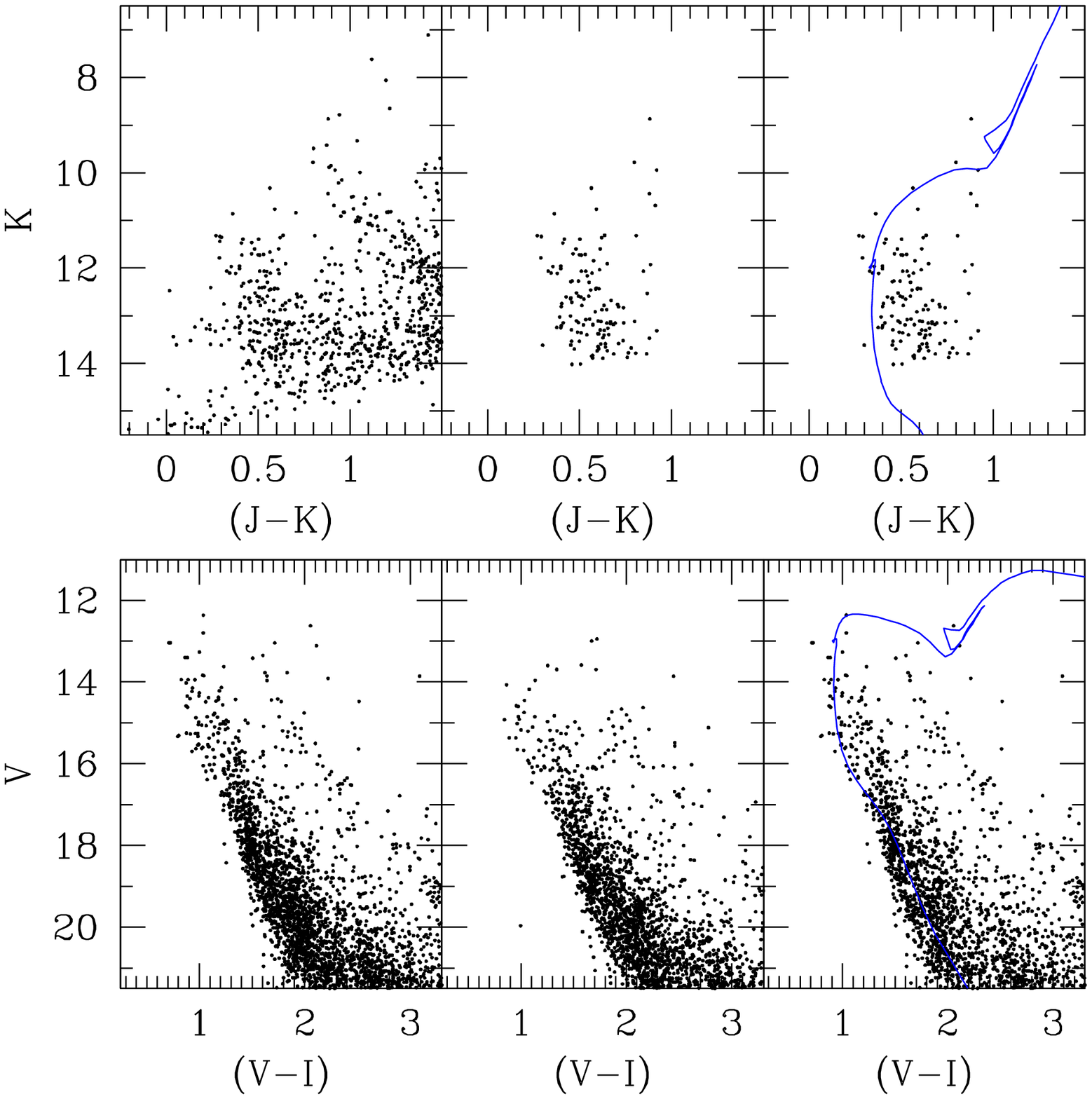}}
\caption{Same as Fig.~8, for Trumpler~34. The blue isochrone is drawn according
to literature parameters, but we do not consider Trumpler~34 as physical group. 
The clean optical CMD contains 2184 stars. See text for more details.}
\end{figure}

\section{Conclusions}

We have presented homogeneous $V,I$ CCD photometry in the field of nine Galactic open clusters,
obtained with the purpose of estimating, in many cases for the first time, their fundamental
parameters. In most cases, this is the first CCD study in the cluster region.

\noindent
We have performed a star count analysis of these fields to assess the clusters'
reality as over-densities of stars with respect to the field, and to estimate their radii.
By means of comparison fields, and applying a statistical subtraction procedure, we have
constructed field star decontaminated CMDs for these clusters. We complemented this data-set
with photometry from 2MASS archive to test and strenghten our findings.

\noindent
The analysis of the optical and IR CMDs,
together with the results from the star counts, allowed us to determine estimates star clusters'
basic parameters.\\

\noindent
Our finding can be summarized as follows:

\begin{description}

\item $\bullet$ all clusters are found to be real, and of intermediate or old age;

\item $\bullet$  Hogg~19 is the oldest cluster of the sample, with an age around 2.5 Gyr; the existence
    of such an old cluster in a hostile environment as the inner Galaxy is puzzling;

\item $\bullet$  Lynga~4 is the most heavily reddened cluster in the sample, and we could detect
   it only in the IR;

\item $\bullet$ Trumpler~20 has been found to be quite an interesting cluster, much similar to
         NGC~7789. The most interesting result is the presence
         of a double red clump, which deserves further investigation.

\end{description}

\noindent
This investigation emphasizes the difficulty to study the inner regions of the Galaxy in the
mere optical domain.  We show, however, that the combination of star counts and CMDs
in the optical and IR, with common
knowledge of the spiral structure of our galaxy, is quite an effective strategy to distinguish
real star clusters from over-densities produced by the patchy distribution of dust, gas and
stars in spiral arms.\\ 

\noindent
Present and future wide area surveys in the IR, conducted by UKIDSS (Lawrence et al. 2007)
and VISTA (McPherson et al. 2004) consortia, will certainly provide more suitable data to discover
and study new star clusters in the inner Galaxy.\\

\section*{acknowledgements}
AFS acknowledges ESO for supporting a visit to Chile through Director General Discretionary
Fundings (DGDF), where this project was completed. EC acknowledges the Chilean Centro de
Astrof\'isica FONDAP (No. 15010003).
The authors are much obliged for the use of the NASA Astrophysics Data System, of the
$SIMBAD$ database (Centre de Donn\'es Stellaires --- Strasbourg, France) and of the WEBDA open cluster 
database. This publication also made use of data from the Two Micron All Sky Survey, 
which is a joint project of the University of Massachusetts and the Infrared 
Processing and Analysis Center/California Institute of Technology, funded by the 
National Aeronautics and Space Administration and the National Science Foundation.

%--------------------------------- References -------------

\end{document}